\documentclass[longauth]{aa} % for the lon lists of affiliations
\usepackage{graphicx}
\usepackage{color}
\usepackage{txfonts}
\newcommand{\lsun}{log$L/L_{\odot}\,$}
\newcommand{\msun}{$M_{\odot}\,$}

\begin{document}

\title{RR Lyrae stars in Galactic globular clusters. VI. The Period-Amplitude relation.}

\author{G. Bono \inst{1,2}, F. Caputo \inst{1}, M. Di Criscienzo \inst{1,3}}

\authorrunning{Bono et al.}

\offprints{G. Bono\\ \email{bono@mporzio.astro.it} }

\institute {INAF-Osservatorio Astronomico di Roma, Via Frascati 33,
00040 Monte Porzio Catone, Italy {\tt bono@mporzio.astro.it}, 
{\tt caputo@mporzio.astro.it}, {\tt dicriscienzo@mporzio.astro.it}\\
\and European Southern Observatory, Karl-Schwarzschild-Str. 2,
D-85748 Garching bei Munchen, Germany\\
\and INAF-Osservatorio Astronomico di Capodimonte, Via Moiariello 16,
80131 Napoli, Italy
 }

\date{Received, accepted}

\abstract
  % context heading (optional)
{}
  % aims heading (mandatory)
{This work uses nonlinear convective models of RR Lyrae stars and 
evolutionary predictions of low-mass helium burning stellar structures to  
constrain the properties of cluster and field RR Lyrae variables. In particular, 
we address two problems: is the Period-Amplitude ($PA_V$) plane of fundamental 
(RR$_{ab}$) variables a good diagnostic for the metal abundance?  
Is the $M_V$(RR)-[Fe/H] relation of field and cluster variables linear over 
the whole metal abundance range of [Fe/H]$\sim -$2.5 to $\sim$ 0?
}
% methods heading (mandatory)
{We perform a detailed comparison between theory and observations for 
fundamental RR Lyrae variables in the solar neighborhood and in both 
Oosterhoff type I (OoI) and type II (OoII) Galactic globular clusters.  
}
% results heading (mandatory)
{We show that the distribution of cluster RR$_{ab}$ variables in the 
$PA_V$ plane depends not only on the metal abundance, but also on the 
cluster Horizontal Branch (HB) morphology. We find that on average the 
observed pulsation parameter $k_{puls}$, connecting the period to the visual 
amplitude, increases when moving from metal-poor to metal-rich GGCs.  
However, this parameter shows marginal changes among OoI clusters 
with intermediate to red HB types and iron abundances $-1.8\le$[Fe/H]$\le-1.1$, 
whereas its value decreases in OoII clusters with the bluer HB morphology,  
although these clusters are also the less metal-poor ones of the group. 
Moreover, at [Fe/H]=$-1.7\pm$0.1 the OoI clusters present redder 
HB types and larger $\langle k_{puls}\rangle$ values than the OoII clusters.
The RR$_{ab}$ variables in $\omega$~Cen and in the solar neighborhood 
further support the evidence that the spread in [Fe/H], at fixed $k_{puls}$, 
is of the order of $\pm$0.5 dex. Using the results of synthetic HB simulations, 
we show that the $PA_V$ plane  can provide accurate cluster distance estimates. 
We find that the RR$_{ab}$ variables in OoI and in OoII clusters with very blue 
HB types obey a well-defined 
$M_V$(RR)-$k_{puls}$ relation, while those in OoII clusters with moderately 
blue HB types present a zero-point that is $\sim 0.05$ mag brighter. Regarding 
field variables, we show that with [Fe/H]$\ge -$1.0 a unique $M_V$(RR)-$k_{puls}$ 
relation can be adopted, independently of the color distribution of the parent HB 
star population.}
% conclusions heading (optional)
{
Current findings suggest that the $PA_V$ distribution  does 
not seem to be a robust diagnostic for the metal abundance of RR$_{ab}$ variables.
However, the same observables can be used to estimate the absolute magnitude of 
globular cluster and field RR$_{ab}$ variables. Eventually, we show that over 
the  metallicity range $-2.4\le[Fe/H]\le 0.0$ the $M_V$(RR)-[Fe/H] relation is 
not linear but has a parabolic behavior.
}


\maketitle
%
%________________________________________________________________

%%%%%%%%%%%%%%%%%%%%%%%%%%%%%%%%%%%%%%%%%%%%%%%%%%%%%%%%%%%%%%%%%%%%%%%%%%%
\section{Introduction}

It has been far-back recognized that the properties of RR Lyrae
variables provide firm constraints to investigate several important
aspects of stellar evolution and cosmology. The calibration of the
absolute visual magnitude $M_V$(RR) as a function of the
iron-to-hydrogen content [Fe/H] is generally used for distance
determinations in the Local Group and the RR Lyrae-based distances 
provide an independent test for the Cepheid distance scale in nearby
galaxies (Magellanic Clouds, M31, dwarf spheroidal galaxies) and for
the calibration of secondary distance indicators such as the
globular cluster luminosity function in more distant galaxies (see
e.g. Di Criscienzo et al. 2006, and references therein). Moreover,
the distance of RR Lyrae stars observed in globular clusters is a
fundamental step to determine the absolute magnitude of the cluster
main-sequence turn-off, which is the classical "clock" to estimate
the age of these ancient stellar systems.

Together with this traditional role for distance determinations, since 
the pioneering investigation by Preston (1959) it has also been suggested 
that the location of fundamental mode variables (RR$_{ab}$)
in the Period-Amplitude ($PA_V$) plane, i.e., in the so-called Bailey diagram, 
depends on the metal abundance. Among the more recent papers, 
we mention Alcock et al. (2000) who used the visual amplitude
of RR$_{ab}$ stars in the globular clusters M15 ([Fe/H]=$-$2.1),
M3 ([Fe/H]=$-$1.6) and M5 ([Fe/H]=$-$1.4) to get the 
calibration

$$[Fe/H]_A=-2.60-8.85\log P_{ab}-1.33A_V\eqno(1)$$

\noindent and Sandage (2004) who determined 

$$[Fe/H]_S=-2.15-7.99\log P_{ab}-1.45A_V\eqno(2)$$

\noindent from field variables with spectroscopic [Fe/H] measurements. Although these 
Period-Metallicity-Amplitude relations 
present a large intrinsic indeterminacy of $\sim$ 0.35 dex, they were 
used by Alcock et al. (2000) to estimate a median metal content 
of [Fe/H]$\sim -$1.6 for a huge
sample of RR$_{ab}$ stars in the bar of the Large Magellanic Cloud
(LMC), by Brown et al. (2004) to derive a mean metallicity of
[Fe/H]=$-1.8\pm$0.3 for the 29 RR$_{ab}$ variables they identified
in a halo field of M31, and by Kinemuchi et al. (2006) to study 
the properties of RR Lyrae stars in the solar neighborhood. 

The suggested dependence of the Bailey diagram on the metal
abundance accounts for the observational evidence that  RR$_{ab}$
stars in Oosterhoff type II globular clusters tend to have, for a
given amplitude, longer periods than those in Oosterhoff type I
clusters. Let us recall that according to the average period 
$\langle P_{ab}\rangle$ of their $ab$-type variables, the globular
clusters are conventionally classified into two Oosterhoff
groups: the former group (Oosterhoff type I: OoI) includes
metal-intermediate clusters with $\langle P_{ab}\rangle\sim$ 0.55
days, while the latter (Oosterhoff type II: OoII) includes
metal-poor clusters with $\langle P_{ab}\rangle\sim$ 0.65 days.
However, one cannot neglect that OoII
clusters show bluer horizontal branch (HB) star distributions than
OoI clusters. Therefore the $PA_V$ diagram, as already suggested
by Clement \& Shelton (1999), migth not depend on the metal
abundance but on the evolutionary status of RR Lyrae stars.

From a theoretical point of view, it is widely accepted that the
pulsation period $P$ is physically governed by the von Ritter relation
$P\rho^{1/2}=Q$ ($\rho$ is the stellar density and $Q$ the pulsation
constant) which yields that the pulsation period is a function of the
pulsator mass $M$, luminosity $L$, and effective temperature $T_e$.
On this ground, since the earlier linear and adiabatic pulsation models,
the $P=f(M,L,T_e)$ relation, the so-called van Albada \& Baker
(1971, 1973) relation, has been at the basis of several
investigations focussed on the estimate of RR Lyrae mass and
luminosity. However, accurate predictions concerning the
luminosity and the radial velocity variations along the pulsation cycle,
and their dependence on the pulsator structural parameters, become
available only with the modern nonlinear, convective approach
(Stellingwerf 1984).

The purpose of the present investigation is to use the updated
and detailed sets of nonlinear, convective models for fundamental (F)
pulsators computed by our group (see Marconi
et al. 2003 [Paper II]; Di Criscienzo, Marconi \& Caputo 2004 [Paper
III], and references therein) to investigate the $PA_V$ relation for
RR$_{ab}$ variables. The theoretical scenario is
discussed in Section 2, while Section 3 deals with the comparison 
with observations. The role of the Period-Amplitude diagram in the distance 
estimate of RR$_{ab}$ variables is presented in Section 4 and 
the conclusions close the paper.

%%%%%%%%%%%%%%%%%%%%%%%%%%%%%%%%%%%%%%%%%%%%%%%%%%%%%%%%%%%%%%%%%%%%%%
\section{The physical meaning of the $PA_V$ relation}

The pulsation models used in the present paper have been computed
with the nonlinear convective, hydrodynamical code which has already
been described in previous investigations (see Paper II, Paper
III, and references therein) and it will not be further discussed.
We only wish to mention that the grid of models covers a wide range
in stellar mass, luminosity, and chemical composition (see Table 1)
and that the bolometric light curves of the models have been
transformed into the observational plane by adopting the bolometric
corrections and color-temperature transformations provided by
Castelli et al. (1997a,b). This approach allows us to derive
light-curve amplitudes $A_i$ and mean absolute magnitudes, either
intensity-weighted $\langle M_i\rangle$ or magnitude-weighted
$(M_i)$, for the various photometric bands.

\begin{table}
\begin{center}
\caption{Main parameters of the pulsation models used in this paper.} \label{PLCM}
\begin{tabular}{lccl}
\hline \hline
$Y$  &    $Z$ & $M$/\msun & \lsun\\
\hline
0.24&    0.0001&  0.80&    1.72, 1.81, 1.91\\
    &          &  0.75&    1.61, 1.72, 1.81\\
    &          &  0.70&    1.72\\
    &          &  0.65&    1.61\\
0.24&    0.0004&  0.70&    1.61, 1.72, 1.81\\
0.24&    0.001 &  0.75&    1.71\\
    &           &  0.65&   1.51, 1.61, 1.72\\
0.255&   0.006 &  0.58&    1.55, 1.65, 1.75\\
\hline
\end{tabular}
\end{center}
\end{table}

The entire set of models pulsating in the fundamental mode shows a
linear correlation between the bolometric amplitude and the
pulsation period (logarithmic scale) in the sense that the
amplitude decreases from short to
long periods, at fixed mass and luminosity. 
Moreover, we found that the luminosity amplitude, at
fixed period, increases as the stellar luminosity increases or as
the stellar mass decreases, but to a lesser extent (see Fig. 3 in
Paper II). In this context, it is worth mentioning that the pulsation limit cycle
stability is also governed by the efficiency of convection as flux
carrier in the stellar envelope, and in turn on the value of the
mixing-length parameter $l/H_p$ adopted to close the system of
convective and hydrodynamical equations. Note that the depth of the
convective region increases when moving from higher to lower effective
temperatures and that convection is the physical mechanism that
quenches pulsation instability. As a consequence, the RR Lyrae
models at constant stellar mass and luminosity show that an
increase in the mixing-length parameter from $l/H_p$=1.5 to 2.0
causes a systematic decrease ($\sim 100$ K) in the effective
temperature of the first overtone blue edge (FOBE) and the simultaneous increase
in the effective temperature of both the blue edge (FBE, $\sim$ 100 K)
and the red edge (FRE, $\sim$ 300 K) of fundamental pulsation. 
As a whole, the increase in the
efficiency of the convective transport causes a narrowing of the width
in temperature of the instability strip. 
On the other hand, the amplitude of fundamental pulsators reaches its maximum value
close to the FBE and attains vanishing values close to the FRE. This yields 
that different assumptions concerning the mixing-length parameter
affect the region of the instability strip where fundamental pulsators
are pulsationally unstable, and in turn both the zero-point and the
slope of the predicted Period-Amplitude relation.

Using the intensity-averaged
$\langle M_V\rangle$ magnitudes of fundamental pulsators with 
$Z$=0.0001-0.006, we find that the correlation
between pulsation period, visual amplitude, magnitude, and mass (in solar units) 
is given by
$$\log P_{ab}=0.136-0.189A_V-0.385\langle M_V\rangle-0.30\log M\eqno(3)$$
for $l/H_p$=1.5, and
$$\log P_{ab}=0.027-0.142A_V-0.385\langle M_V\rangle-0.35\log M\eqno(4)$$
for $l/H_p$=2.0, where the rms dispersion of the fit 
is 0.025 dex. For the sake of the following discussion, let us emphasize that in these 
relations the pulsator mass and luminosity are free parameters. 

According to these relations, the RR$_{ab}$ distribution in the $PA_V$
diagram is described by the {\it pulsation} parameter 

$$k(1.5)_{puls}=0.136-\log P_{ab}-0.189A_V$$
or
$$k(2.0)_{puls}=0.027-\log P_{ab}-0.142A_V,$$

\noindent which in turn depends on the pulsator {\it evolutionary}
properties as 

$$k(1.5)_{ev}=0.385\langle M_V\rangle+0.30\log M$$
and
$$k(2.0)_{ev}=0.385\langle M_V\rangle+0.35\log M$$

%%%%%%%%%%%%%%%%%%%%%%%%%%%%%%%%%%%%%%%%%%%%%%%%%%%%%%%%%%%%%%%%%%%%%%%

\begin{figure}
\begin{center}
\includegraphics[width=8cm]{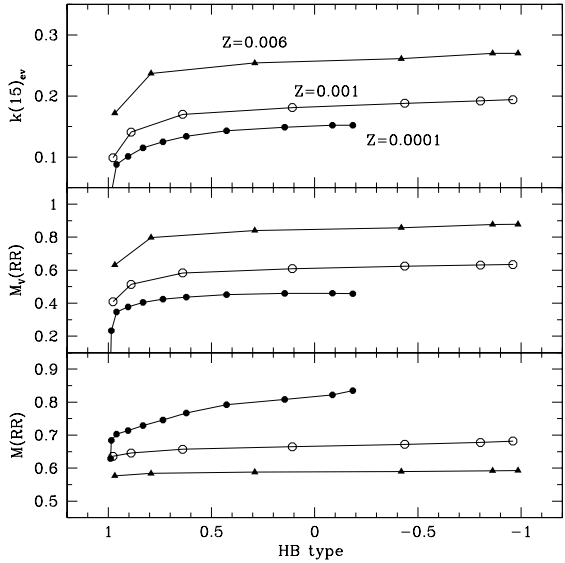}
\caption{From bottom to top: the average mass $M$(RR) in solar units,
the absolute visual magnitude $M_V$(RR), and the evolutionary parameter
$k(1.5)_{ev}$ as a function of the HB type. Current predictions rely on a
set of SHB simulations discussed in Paper IV.}
\end{center}
\end{figure}

At variance with the pulsational parameters 
$k(1.5)_{puls}$ and $k(2.0)_{puls}$, the values of the evolutionary  
ones $k(1.5)_{ev}$ and
$k(2.0)_{ev}$ cannot be directly estimated
from observations. However, all the synthetic horizontal branches (SHB)
simulations (see, e.g., Demarque et al. 2000; Catelan et al. 2004;
Cassisi et al. 2004, [Paper IV]) agree in suggesting that, for a fixed metallicity,
the average mass of HB stars in the RR Lyrae region decreases when
moving from red to blue HB morphologies, whereas the average
luminosity presents an opposite trend. Furthermore, for a
fixed HB morphology, an increase in the metal content causes a
decrease in the same intrinsic parameters. 

\begin{table*}
\begin{center}
\caption{Selected results of SHB simulations with $Z\le$ 0.006. For each metal content $Z$
and mean mass of HB stars $M$(HB),
we list the predicted mean values of the
HB type and of the RR Lyrae mass, absolute magnitude and $k_{ev}$ parameters,
together with the rms dispersion about the mean. The masses are in solar units.}\label{PLCM}
\begin{tabular}{lccccc}
\hline \hline
$M$(HB) &$\langle$HB$\rangle$&$\langle M$(RR)$\rangle$ & $\langle M_V$(RR)$\rangle$ &   $\langle k(1.5)_{ev}\rangle$            &   $\langle k(2.0)_{ev}\rangle$            \\
\hline
\multicolumn{6}{c}{$Z$=0.0001, $Y$=0.23}\\
0.68    &   +0.96$\pm$0.01    &   0.70$\pm$0.02 &   0.35$\pm$0.02 &   0.088$\pm$0.012 &   0.080$\pm$0.012 \\
0.70    &   +0.90$\pm$0.02    &   0.71$\pm$0.03 &   0.38$\pm$0.01 &   0.101$\pm$0.006 &   0.094$\pm$0.006 \\
0.72    &   +0.83$\pm$0.02    &   0.73$\pm$0.03 &   0.41$\pm$0.01 &   0.115$\pm$0.004 &   0.108$\pm$0.006 \\
0.74    &   +0.74$\pm$0.02    &   0.75$\pm$0.03 &   0.42$\pm$0.01 &   0.125$\pm$0.002 &   0.119$\pm$0.002 \\
0.76    &   +0.62$\pm$0.03    &   0.77$\pm$0.03 &   0.44$\pm$0.01 &   0.134$\pm$0.002 &   0.128$\pm$0.002 \\
0.78    &   +0.43$\pm$0.03    &   0.79$\pm$0.03 &   0.45$\pm$0.01 &   0.143$\pm$0.002 &   0.138$\pm$0.002 \\
0.80    &   +0.15$\pm$0.01    &   0.81$\pm$0.03 &   0.46$\pm$0.01 &   0.149$\pm$0.002 &   0.144$\pm$0.002 \\
0.82    &   $-$0.09$\pm$0.02  &0.82$\pm$0.02  &     0.46$\pm$0.01 &   0.152$\pm$0.002 &   0.148$\pm$0.002 \\
0.84    &   $-$0.19$\pm$0.02  &0.84$\pm$0.02    &   0.46$\pm$0.01 &   0.152$\pm$0.002 &   0.148$\pm$0.002 \\
\multicolumn{6}{c}{$Z$=0.0003, $Y$=0.23}\\

0.64    &   +0.97$\pm$0.01  &   0.66$\pm$0.02 &   0.36$\pm$0.04 &   0.084$\pm$0.030 &   0.075$\pm$0.030 \\
0.66    &   +0.91$\pm$0.01  &   0.68$\pm$0.02 &   0.44$\pm$0.02 &   0.117$\pm$0.008 &   0.109$\pm$0.008 \\
0.68    &   +0.78$\pm$0.01  &   0.70$\pm$0.02 &   0.49$\pm$0.01 &   0.142$\pm$0.004 &   0.134$\pm$0.004 \\
0.70    &   +0.52$\pm$0.02  &   0.71$\pm$0.02 &   0.53$\pm$0.01 &   0.161$\pm$0.002 &   0.154$\pm$0.002 \\
0.72    &   +0.11$\pm$0.03  &   0.73$\pm$0.02 &   0.55$\pm$0.01 &   0.170$\pm$0.002 &   0.163$\pm$0.002 \\
0.74    &   $-$0.29$\pm$0.03  &0.74$\pm$0.02    &   0.56$\pm$0.01 &   0.174$\pm$0.002 &   0.168$\pm$0.002 \\
0.76    &   $-$0.60$\pm$0.02  &0.75$\pm$0.02    &   0.56$\pm$0.01 &   0.176$\pm$0.002 &   0.170$\pm$0.002 \\
0.78    &   $-$0.82$\pm$0.01  &0.76$\pm$0.02    &   0.55$\pm$0.01 &   0.176$\pm$0.002 &   0.170$\pm$0.002 \\
\multicolumn{6}{c}{$Z$=0.0006, $Y$=0.23}\\
0.62    &   +0.97$\pm$0.01  &   0.65$\pm$0.02 &   0.37$\pm$0.03 &   0.087$\pm$0.026 &   0.078$\pm$0.026 \\
0.64    &   +0.89$\pm$0.01  &   0.66$\pm$0.02 &   0.48$\pm$0.02 &   0.131$\pm$0.012 &   0.122$\pm$0.012 \\
0.66    &   +0.70$\pm$0.02  &   0.68$\pm$0.02 &   0.53$\pm$0.01 &   0.155$\pm$0.006 &   0.147$\pm$0.006 \\
0.68    &   +0.28$\pm$0.03  &   0.69$\pm$0.02 &   0.57$\pm$0.01 &   0.170$\pm$0.002 &   0.162$\pm$0.002 \\
0.70    &   $-$0.24$\pm$0.02  &0.70$\pm$0.02    &   0.58$\pm$0.01 &   0.176$\pm$0.002 &   0.168$\pm$0.002 \\
0.72    &   $-$0.69$\pm$0.02  &0.70$\pm$0.02    &   0.59$\pm$0.01 &   0.180$\pm$0.002 &   0.172$\pm$0.002 \\
0.74    &   $-$0.91$\pm$0.01  &0.71$\pm$0.02    &   0.59$\pm$0.01 &   0.182$\pm$0.002 &   0.174$\pm$0.002 \\
0.76    &   $-$0.98$\pm$0.01  &0.72$\pm$0.02    &   0.59$\pm$0.01 &   0.183$\pm$0.002 &   0.176$\pm$0.002 \\
\multicolumn{6}{c}{$Z$=0.001, $Y$=0.23}\\
0.60    &   +0.98$\pm$0.01    &   0.64$\pm$0.01 &   0.41$\pm$0.06 &   0.099   $\pm$0.040  &   0.089$\pm$0.040 \\
0.62    &   +0.89$\pm$0.01    &   0.65$\pm$0.01 &   0.51$\pm$0.03 &   0.141   $\pm$0.018  &   0.132$\pm$0.018 \\
0.64    &   +0.64$\pm$0.02    &   0.66$\pm$0.01 &   0.58$\pm$0.02 &   0.170   $\pm$0.008  &   0.161$\pm$0.008 \\
0.66    &   +0.11$\pm$0.02    &   0.67$\pm$0.01 &   0.61$\pm$0.01 &   0.181   $\pm$0.002  &   0.172$\pm$0.002 \\
0.68    &   $-$0.44$\pm$0.03  &0.67$\pm$0.01    &   0.62$\pm$0.01 &   0.188   $\pm$0.002  &   0.180$\pm$0.002 \\
0.70    &   $-$0.80$\pm$0.02  &0.68$\pm$0.01    &   0.63$\pm$0.01 &   0.192   $\pm$0.002  &   0.184$\pm$0.002 \\
0.72    &   $-$0.96$\pm$0.01  &0.68$\pm$0.01    &   0.63$\pm$0.01 &   0.194   $\pm$0.002  &   0.186$\pm$0.002 \\
\multicolumn{6}{c}{$Z$=0.003, $Y$=0.23}\\
0.56    &   +0.97$\pm$0.01    &   0.60$\pm$0.01 &   0.56$\pm$0.07 &   0.149   $\pm$0.040  &   0.138$\pm$0.040 \\
0.58    &   +0.84$\pm$0.02    &   0.61$\pm$0.01 &   0.68$\pm$0.03 &   0.198   $\pm$0.020  &   0.187$\pm$0.020 \\
0.60    &   +0.40$\pm$0.03    &   0.61$\pm$0.01 &   0.73$\pm$0.01 &   0.218   $\pm$0.004  &   0.208$\pm$0.004 \\
0.62    &   $-$0.25$\pm$0.03  &0.62$\pm$0.01    &   0.76$\pm$0.01 &   0.227   $\pm$0.004  &   0.217$\pm$0.004 \\
0.64    &   $-$0.77$\pm$0.02  &0.62$\pm$0.01    &   0.77$\pm$0.01 &   0.234   $\pm$0.006  &   0.223$\pm$0.006 \\
0.66    &   $-$0.97$\pm$0.01  &0.62$\pm$0.01    &   0.78$\pm$0.01 &   0.236   $\pm$0.008  &   0.226$\pm$0.008 \\
\multicolumn{6}{c}{$Z$=0.006, $Y$=0.245}\\
0.54    &   +0.97$\pm$0.01    &   0.58$\pm$0.01 &   0.63$\pm$0.10 &   0.172   $\pm$0.070  &   0.160   $\pm$0.070  \\
0.56    &   +0.79$\pm$0.02    &   0.58$\pm$0.01 &   0.80$\pm$0.03 &   0.237   $\pm$0.012  &   0.226   $\pm$0.012  \\
0.58    &   +0.29$\pm$0.03    &   0.59$\pm$0.01 &   0.84$\pm$0.01 &   0.254   $\pm$0.008  &   0.243   $\pm$0.008  \\
0.60    &   $-$0.42$\pm$0.02  &0.59$\pm$0.01    &   0.86$\pm$0.01 &   0.261   $\pm$0.004  &   0.250   $\pm$0.004  \\
0.62    &   $-$0.86$\pm$0.01  &0.59$\pm$0.01    &   0.88$\pm$0.01 &   0.270   $\pm$0.006  &   0.258   $\pm$0.006  \\
0.64    &   $-$0.98$\pm$0.01  &0.59$\pm$0.01    &   0.88$\pm$0.03 &   0.270   $\pm$0.032  &   0.259   $\pm$0.032  \\
\hline
\end{tabular}
\end{center}
\end{table*}

Using the SHBs computed in Paper IV for various chemical compositions, 
we show in Table 2 some selected predictions based on SHB simulations 
in which the number of predicted RR Lyrae stars approaches $\sim$ 2\% of 
the global HB star population. For each assumed chemical composition 
and mean mass\footnote{The SHBs have been computed by assuming a gaussian
random distribution of HB masses centered on $M$(HB) and with a standard
deviation $\sigma\sim0.02M_{\odot}$.} $M(HB)$
of HB stars, we give the average HB type\footnote{This parameter is 
the ratio (B$-$R)/(B+V+R) among the
numbers of HB stars to the blue (B), within (V) and to the red (R)
of the RR Lyrae instability strip (Lee 1990).} and 
the predicted mean values of the
RR Lyrae mass, absolute magnitude, and $k_{ev}$ parameter,
together with the rms dispersion about the mean. Note that these 
mean values are derived by averaging the results of 10
different simulations. 

Data listed in Table 2 (see also Fig. 1) disclose four substantial points:
\begin{enumerate}
\item the mass range of the predicted RR Lyrae decreases with increasing 
metal content, when moving from very blue to very red HB type distributions; 
\item the $k_{ev}$ parameter, at fixed metallicity, attains rather constant 
values from red to moderately blue HB morphology (i.e., for HB type ranging 
from $\sim -$0.9 to $\sim$ +0.5), whereas it significantly decreases for 
the bluer populations;
\item the $k_{ev}$ parameter, at constant HB type, increases when moving 
from low to high metal abundances. However, for HB$\ge$ +0.9 the metallicity 
effect tends to vanish;
\item the size of this metallicity effect varies with the metallicity range. 
In particular, for HB type $\sim$ 0 we get $\Delta k(1.5)_{ev}\sim$ 0.03
for $0.0001 \le Z \le 0.001$ and $\Delta k(1.5)_{ev}\sim$ 0.08
for $0.001 \le Z \le 0.006$.
\end{enumerate}

In summary, the constraints on $k_{ev}$ provided by the evolutionary 
predictions suggest that the $PA_V$ distribution of RR$_{ab}$ stars in
globular clusters depends {\it both on the cluster metal abundance 
and on the HB morphology}.

%%%%%%%%%%%%%%%%%%%%%%%%%%%%%%%%%%%%%%%%%%%%%%%%%%%%%%%%%%%%%%%

\section{Observed $PA_V$ diagrams}
\subsection{Galactic globular clusters}

\begin{table}
\begin{center}
\caption{Selected parameters for Galactic globular clusters:
HB type, average period of $ab$-type RR Lyrae stars and
iron-to-hydrogen content [Fe/H]$_K$ according to the
Kraft \& Ivans (2003) metallicity scale. For $\omega$~Cen,
we list the average [Fe/H]$_R$ value from Rey et al. (2000) data.}\label{PLCM}
\begin{tabular}{lccc}
\hline \hline
Name &   HB  &    $\langle$log$P_{ab}\rangle$ &   [Fe/H]$_K$  \\
\hline\\
\multicolumn{4}{c}{Oosterhoff type II}\\
N4590~(M68)   &   +0.44    &   $-$0.201    &   $-$2.43    \\
N6426         &   +0.58    &   $-$0.153    &   $-$2.43    \\
N7078~(M15)   &   +0.67    &   $-$0.189    &   $-$2.42    \\
N5053         &   +0.52    &   $-$0.174    &   $-$2.41    \\
N6341~(M92)   &   +0.91    &   $-$0.195    &   $-$2.38    \\
N5466         &   +0.58    &   $-$0.172    &   $-$2.22   \\
N5024~(M53)   &   +0.81    &   $-$0.189    &   $-$2.02    \\
N6809~(M55)   &   +0.87    &   $-$0.181    &   $-$1.85    \\
N6333~(M9)    &   +0.87    &   $-$0.203    &   $-$1.79   \\
N7089~(M2)    &   +0.96    &   $-$0.168    &   $-$1.56    \\
\hline\\
\multicolumn{4}{c}{Oosterhoff type I}\\
N4147       &   +0.55      &   $-$0.282    &   $-$1.79    \\
I4499      &   +0.11      &   $-$0.238    &   $-$1.60     \\
N6934       &   +0.25      &   $-$0.252    &   $-$1.59    \\
N3201       &   +0.08      &   $-$0.252    &   $-$1.56    \\
N5272~(M3)  &   +0.08      &   $-$0.257    &   $-$1.50   \\
N7006       &   $-$0.28    &   $-$0.246    &   $-$1.48    \\
N6715~(M54) &   +0.75      &   $-$0.237    &   $-$1.47    \\
N6981~(M72) &   +0.14      &   $-$0.256    &   $-$1.42    \\
N6229       &   +0.24      &   $-$0.270    &   $-$1.41  \\
N6864~(M75) &   $-$0.07    &   $-$0.231    &   $-$1.29    \\
N5904~(M5)  &   +0.31      &   $-$0.263    &   $-$1.26    \\
N1851       &   $-$0.36    &   $-$0.241    &   $-$1.19    \\
N6121~(M4)  &   $-$0.06    &   $-$0.275    &   $-$1.15    \\
N6362       &   $-$0.58    &   $-$0.265    &   $-$1.15    \\
N6723       &   $-$0.08    &   $-$0.262    &   $-$1.11    \\
N6171~(M107)&   $-$0.73    &   $-$0.272    &   $-$1.10   \\
\hline\\
\multicolumn{4}{c}{Peculiar clusters}\\
N5139~($\omega$~Cen) &+0.92&   $-$0.189    &   $-$1.62 \\
N6441               &$-$0.73&   $-$0.132   &   $-$0.85  \\
\hline
\end{tabular}
\end{center}
\end{table}

\begin{table}
\begin{center}
\caption{Mean $k(1.5)_{puls}$ and $k(2.0)_{puls}$ values for RR$_{ab}$ stars in Galactic
globular clusters.} \label{PLCM}
\begin{tabular}{lcccc}
\hline \hline
Name& HB & [Fe/H]$_K$ & $\langle k(1.5)_{puls}\rangle$ & $\langle k(2.0)_{puls}\rangle$\\
\hline\\
\multicolumn{5}{c}{Oosterhoff type II}\\
N4590   &   +0.44   &   $-$2.43 &  0.181$\pm$0.026 &   0.118$\pm$0.024 \\
N6426   &   +0.58   &   $-$2.43 &  0.131$\pm$0.018 &   0.068$\pm$0.017 \\
N7078   &   +0.67   &   $-$2.42 &  0.184$\pm$0.030 &   0.118$\pm$0.030 \\
N5053   &   +0.52   &   $-$2.41 &  0.163$\pm$0.045 &   0.095$\pm$0.036 \\
N6341   &   +0.91   &   $-$2.38 &  0.144$\pm$0.026 &   0.089$\pm$0.021 \\
N5466   &   +0.58   &   $-$2.22 &  0.116$\pm$0.041 &   0.063$\pm$0.031 \\
N5024   &   +0.81   &   $-$2.02 &  0.137$\pm$0.029 &   0.080$\pm$0.029 \\
N6809   &   +0.87   &   $-$1.85 &  0.132$\pm$0.024 &   0.077$\pm$0.018 \\
N6333   &   +0.87   &   $-$1.79 &  0.130$\pm$0.017 &   0.080$\pm$0.016 \\
N7089   &   +0.96   &   $-$1.56 &  0.132$\pm$0.031 &   0.073$\pm$0.033 \\
\hline\\
\multicolumn{5}{c}{Oosterhoff type I}\\
N4147   &   +0.55   &   $-$1.79 &  0.215$\pm$0.021 &   0.164$\pm$0.025 \\
I4499   &   +0.11   &   $-$1.60  &  0.209$\pm$0.036 &   0.149$\pm$0.033 \\
N6934   &   +0.25   &   $-$1.59 &  0.219$\pm$0.037 &   0.160$\pm$0.031 \\
N3201   &   +0.08   &   $-$1.56 &  0.211$\pm$0.042 &   0.154$\pm$0.034 \\
N5272   &   +0.08   &   $-$1.50  &  0.194$\pm$0.032 &   0.142$\pm$0.027 \\
N7006   &   $-$0.28 &   $-$1.48 &  0.222$\pm$0.027 &   0.160$\pm$0.023 \\
N6715   &   +0.75   &   $-$1.47 &  0.220$\pm$0.037 &   0.157$\pm$0.032 \\
N6981   &   +0.14   &   $-$1.42 &  0.244$\pm$0.035 &   0.180$\pm$0.033 \\
N6229   &   +0.24   &   $-$1.41 &  0.219$\pm$0.024 &   0.164$\pm$0.019 \\
N6864   &   $-$0.07 &   $-$1.29 &  0.201$\pm$0.059 &   0.141$\pm$0.055 \\
N5904   &   +0.31   &   $-$1.26 &  0.211$\pm$0.049 &   0.156$\pm$0.048 \\
N1851   &   $-$0.36 &   $-$1.19 &  0.198$\pm$0.040 &   0.141$\pm$0.035 \\
N6121   &   $-$0.06 &   $-$1.15 &  0.184$\pm$0.030 &   0.142$\pm$0.034 \\
N6362   &   $-$0.58 &   $-$1.15 &  0.218$\pm$0.030 &   0.162$\pm$0.029 \\
N6723   &   $-$0.08 &   $-$1.11 &  0.210$\pm$0.049 &   0.155$\pm$0.042 \\
N6171   &   $-$0.73 &   $-$1.10  &  0.239$\pm$0.043 &   0.179$\pm$0.041 \\
\hline\\
\multicolumn{5}{c}{Peculiar clusters}\\
N5139   &   +0.92   &   $-$1.62 &   0.150$\pm$0.043 &   0.092$\pm$0.052 \\
N6441   &  $-$0.73  &   $-$0.85 &   0.106$\pm$0.025 &   0.048$\pm$0.025 \\
\hline
\end{tabular}
\end{center}
\end{table}

For the RR Lyrae stars in Galactic globular clusters for which the
visual amplitude $A_V$ is available in the literature, Table 3 gives
the observed HB type (Harris 2003)\footnote{http://physwww.physics.mcmaster.ca/\%7Eharris/mwgc.dat)}, 
the average period of $RR_{ab}$
variables and the iron-to-hydrogen content [Fe/H]$_K$ on the Kraft \& Ivans
(2003) metallicity scale. For $\omega$~Cen, whose RR Lyrae stars are
characterized by a wide spread in metal abundance, we list the
average value ([Fe/H]=$-1.62\pm$0.27) based on Rey et al. (2000)
data and the HB type determined by Piersimoni et al. (2007, in
preparation). As far as NGC~6441 is concerned, the HB type has been 
determined by Catelan (2005) although it should be
mentioned that this cluster shows a very unusual HB extending from a
stubby red to a very blue component (Rich et al. 1997). Moreover,
the periods of the observed RR$_{ab}$ variables are too long for the
current cluster metallicity, thus hampering a safe Oosterhoff 
classification (see e.g. Pritzl et al. 2001).

\begin{figure}
\begin{center}
\includegraphics[width=8cm]{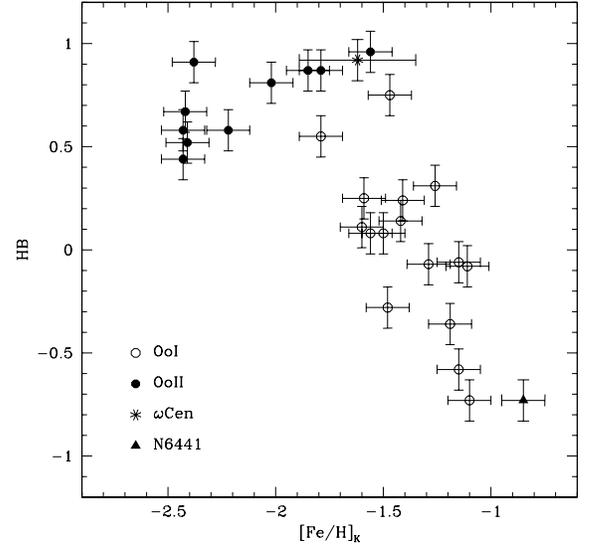}
\caption{The HB type versus the metal content [Fe/H]$_K$ for
Oosterhoff type II (OoII, filled circles) and
Oosterhoff type I (OoI, open circles) Galactic globular clusters.
The error bars have been estimated by assuming $\epsilon (HB)=\pm0.1$
and $\epsilon [Fe/H]_K=\pm0.1$  For $\omega$~Cen (asterisk),
we plot the average value [Fe/H]$_R=-1.62\pm$0.27 derived by
Rey et al. (2000).}
\end{center}
\end{figure}

Figure 2 shows the cluster HB type as a function of the metal
content [Fe/H]$_K$.  It is worth noticing that even the selected
sample of RR Lyrae-rich globular clusters presents the so-called
{\it second parameter} problem: in order to account for the observed
HB morphology, together with the metal abundance it is required a 
further intrinsic parameter.  
However, we also note that OoI and
OoII clusters seem to follow quite different behaviors: the HB
morphology of the OoI clusters becomes bluer as the metal content 
decreases, whereas for the latter group the HB morphology becomes
bluer as the cluster becomes more metal-rich. As a consequence, the
OoII clusters with very blue HB morphology, including $\omega$~Cen,
appear to be the ``natural'' extension of OoI clusters to lower
metal abundances.

\begin{figure}
\begin{center}
\includegraphics[width=8cm]{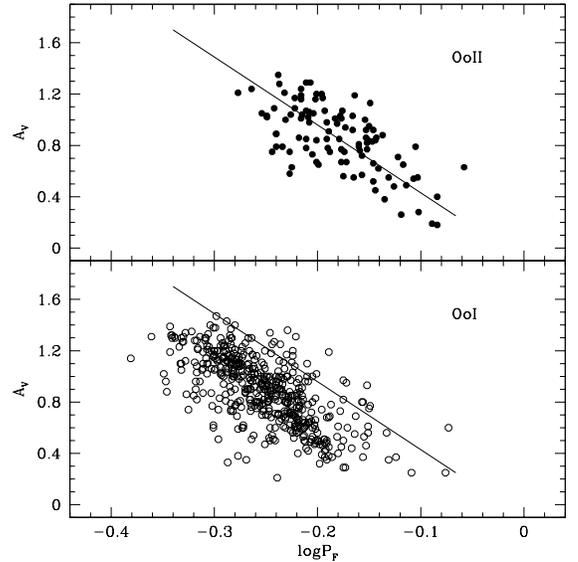}
\caption{Visual amplitude versus period for RR$_{ab}$ stars
in Oosterhoff type II (OoII, top panel) and Oosterhoff type I (OoI, bottom panel)
globular clusters. The solid line shows the ridge line of variables
in OoII clusters and is based on the predicted slope
$\delta$log$P_F/\delta A_V=-$0.189.}
\end{center}
\end{figure}

Figure 3 shows the $PA_V$ diagram of the observed RR$_{ab}$ stars
in OoII (top panel) and OoI (bottom panel) clusters. The variables
in $\omega$ Cen and in NGC~6441 have not been included in this figure
and will be discussed separately. The solid line in the top panel is
the ridge line of variables in OoII clusters and it was drawn by
adopting the predicted slope $\delta$log$P_F/\delta A_V=-$0.189 [see
Eq. (3)]. The same line is also plotted in the bottom panel to
emphasize that RR Lyrae stars in OoI clusters present systematically shorter period, 
at fixed pulsation amplitude.

\begin{figure}
\begin{center}
\includegraphics[width=8cm]{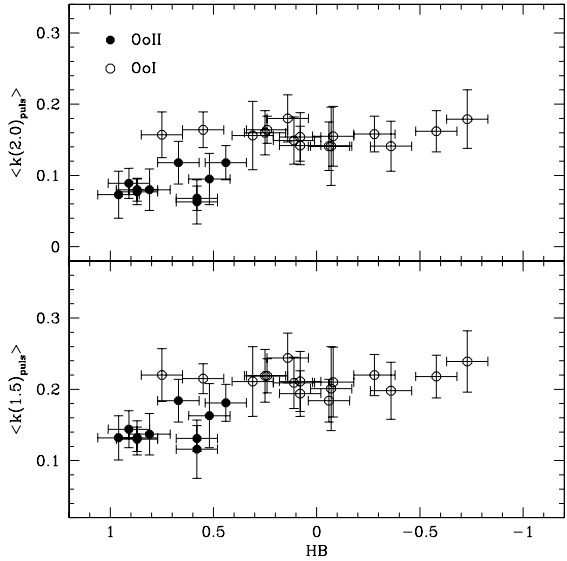}
\caption{The average $\langle k(1.5)_{puls}\rangle$ (bottom panel) and 
$\langle k(2.0)_{puls}\rangle$ (top panel) values for RR$_{ab}$ stars
in Oosterhoff type I (OoI, open circles) and Oosterhoff type II
(OoII, filled circles) globular clusters plotted as a function of the
cluster HB type.}
\end{center}
\end{figure}

\begin{figure}
\begin{center}
\includegraphics[width=8cm]{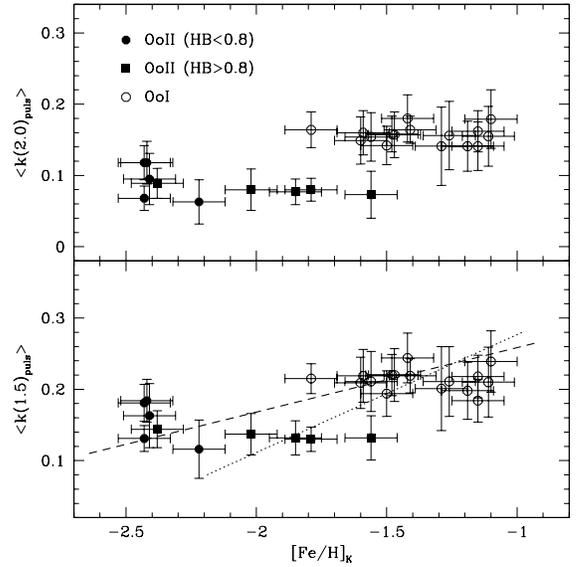}
\caption{The average $\langle k(1.5)_{puls}\rangle$ (bottom panel) and 
$\langle k(2.0)_{puls}\rangle$ (top panel) values for RR$_{ab}$ stars
in Oosterhoff type I (OoI, open circles) and Oosterhoff type II
(OoII) globular clusters as a function of metal abundance. The filled
circles mark OoII clusters with HB type ranging from red to
moderately blue, while the filled squares refer to those with very
blue HB type. The dashed and the dotted lines in the bottom panel
display two different choices in the selection of the calibrating
clusters. See text for more details.}
\end{center}
\end{figure}

Based on the data plotted in Fig. 3, we derive the average $\langle
k(1.5)_{puls}\rangle$ and $\langle k(2.0)_{puls}\rangle$ values
listed in Table 4 together with the their standard deviations. 
Figures 4 and 5 show these parameters versus the cluster HB type 
and the metal content [Fe/H]$_K$, respectively. In the latter figure, 
the OoII clusters are also selected according to the HB morphology.  

As a whole, we find that:
\begin{itemize}
\item among the OoI clusters, which have metal abundances 
from [Fe/H]=$-$1.8 to $-$1.1 and HB types redder than +0.75,
neither the $\langle k(1.5)_{puls}\rangle$ nor the 
$\langle k(2.0)_{puls}\rangle$ parameter show significant variations
with [Fe/H] or HB type;
\item among the OoII clusters, which have metal abundances ranging 
from [Fe/H]=$-$2.4 to $-$1.6 and HB types bluer than +0.44, both
$\langle k(1.5)_{puls}\rangle$ and $\langle k(2.0)_{puls}\rangle$
present a mild decrease for the clusters with bluer HB morphologies,
though they are also the less metal-poor ones of the group;
\item for [Fe/H]=$-1.7\pm$0.1, where both OoI and OoII clusters
are observed, the former clusters have redder HB types {\it and}
larger $\langle k_{puls}\rangle$ values than the latter ones.
\end{itemize}

Bearing in mind the above discussion on the $k_{ev}$
values listed in Table 2, the Bailey diagram of the RR$_{ab}$ stars
observed in Galactic globular clusters agrees with the evolutionary
prescriptions and does not support the use of a unique $PA_V$ relation
for robust metal abundance determinations. The linear fit over the 
entire dataset plotted in the bottom panel of Fig. 5 gives 
[Fe/H]$\sim -3.1+7.7k(1.5)_{puls}$. This relation 
would predict the RR$_{ab}$ metallicity with the unpleasant average 
uncertainty of $\sim$ 0.4 dex. The intrinsic error becomes even 
worse if the adopted empirical calibration relies on individual
clusters. The use of OoI clusters together with
OoII clusters with moderately blue HB morphology yields 
[Fe/H]$\sim -3.9+11.1k(1.5)_{puls}$ (see the dashed line in the 
bottom panel of Fig. 5), while the use of OoI clusters together 
with OoII clusters with very blue HB morphology yields (see 
the dotted line) [Fe/H]$\sim -2.7+5.8k(1.5)_{puls}$.
Note that the application of the former relation to RR$_{ab}$
variables in OoII clusters with very blue HB stellar populations
would underestimate by $\sim$ 0.7 dex the metallicity of these
variables, while the application of the latter relation to
RR$_{ab}$ variables in OoII clusters with moderately blue 
HB stellar populations would overestimate by $\sim$ 0.5 dex 
the metallicity of the these variables.

\subsection{NGC 6441 and $\omega$ Cen}

Figure 6 shows the $PA_V$ diagram of $ab$-type variables
in NGC~6441 and $\omega$~Cen together with the ridge line
of Oo II variables (see Fig. 3). Data plotted in this
figure support the evidence that all
the RR$_{ab}$ stars in NGC~6441 behave as OoII variables
(see also the $\langle k(1.5)_{puls}\rangle$ and $\langle
k(2.0)_{puls}\rangle$ values listed in Table 4) suggesting
that the RR Lyrae metal abundance is significantly lower than the
current cluster value. This is at odds with the recent spectroscopic
measurements by Clementini et al. (2005) confirming that the 
RR Lyrae stars in NGC~6441 are metal-rich with [Fe/H]$\sim -0.7\pm 0.3$, 
on the Zinn \& West (1984) scale (see also Gratton et al. 2007, and 
references therein) . 

\begin{figure}
\begin{center}
\includegraphics[width=8cm]{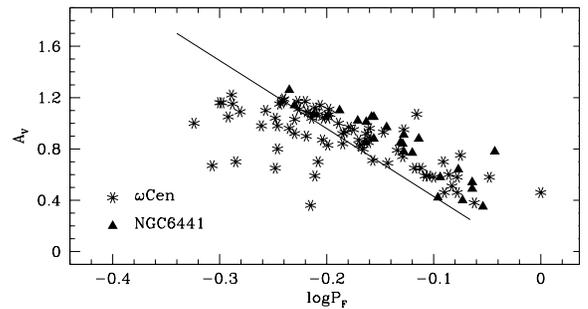}
\caption{Same as in Fig. 3, but for RR$_{ab}$ stars in the two peculiar 
clusters NGC~6441 (triangles) and $\omega$~Cen (asterisks).}
\end{center}
\end{figure}

On the other hand, if the NGC~6441 variables are
generated by the very blue HB component, we should expect small
$k_{ev}$ values even with large metal abundances. However, even 
adopting HB$\sim$ +0.97 the SHB simulations for $Z$=0.003 and $Y\sim$ 0.25 
presented in Table 2 suggest $\langle k(1.5)_{ev}\rangle \sim$ 0.15 and
$\langle k(2.0)_{ev}\rangle\sim$ 0.14 which are
larger than the observed values. Since $k_{ev}$ significantly depends  
on the pulsator luminosity, this discrepancy might imply a larger helium 
content, as recently suggested by Caloi \& D'Antona (2007) who give 
$Y\sim$ 0.37. However, it is worth mentioning that star counts of HB and 
red giant branch stars in  NGC~6441 provided by Layden et al. (1999) do 
not support the high helium abundance scenario. The new HB simulations 
with $Y$=0.30 (Caputo et al. 2007, in preparation) and
the modeling of the observed light curves (Clementini \& Marconi 2007,
in preparation) will probably shed new lights on the unusual properties
of the NGC~6441 RR Lyrae variables.

\begin{figure}
\begin{center}
\includegraphics[width=8cm]{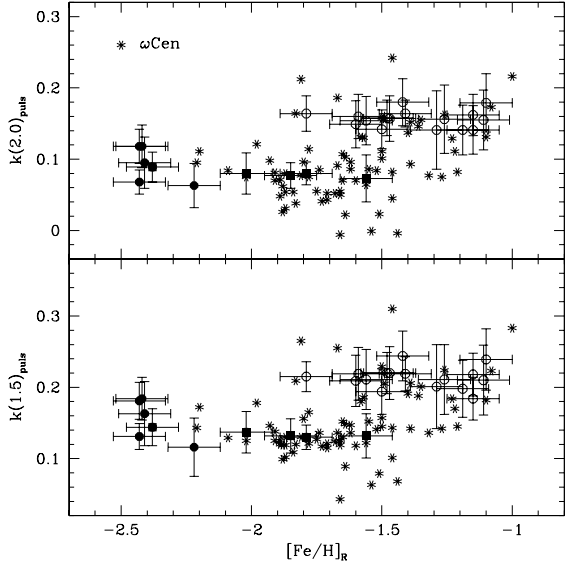}
\caption{The $k(1.5)_{puls}$ (bottom panel) and the $k(2.0)_{puls}$ (top panel) 
parameter for RR$_{ab}$ stars in $\omega$~Cen (asterisks) as a function of 
the metal abundance. Open circles, filled circles, and filled squares display 
Galactic globular clusters and have the same meaning as in Fig. 5.}
\end{center}
\end{figure}

\begin{figure}
\begin{center}
\includegraphics[width=8cm]{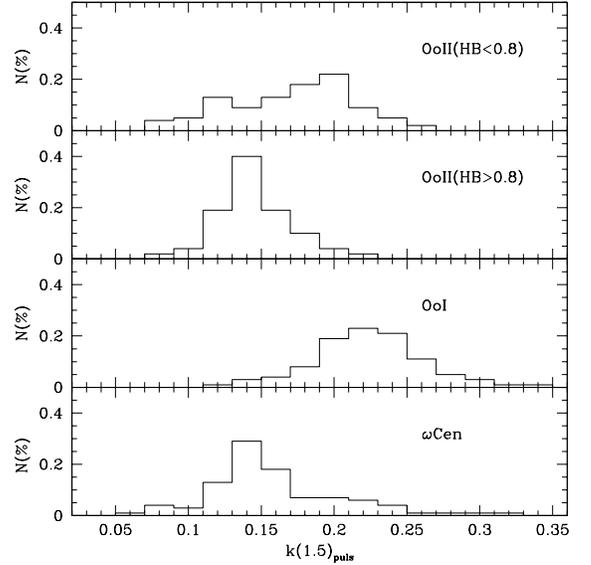}
\caption{Distribution of the $k(1.5)_{puls}$ parameter for RR$_{ab}$ stars 
in $\omega$~Cen (bottom panel) together with OoI and OoII Galactic globular clusters.}
\end{center}
\end{figure}

Regarding the variables in $\omega$~Cen, we plot in
Fig. 7 the $k(1.5)_{puls}$ and $k(2.0)_{puls}$ values versus the
[Fe/H]$_R$ metal abundance determined by Rey et al. (2000). We note
again the quite large dispersion of the metallicity at constant
$k_{puls}$, thus stressing once more the misleading use of the
Bailey diagram for reliable metal abundance determinations. The
comparison with the Galactic globular cluster data presented in Fig.
5, here repeated for the sake of clearness, indicates that the bulk
of RR$_{ab}$ stars in $\omega$~Cen behave as the variables in OoII
clusters, with a minor fraction sharing the properties of the OoI
variables (see also Clement \& Rowe 2000). However, it is worth
noticing that, consistently with the $\omega$~Cen HB type, the
agreement with the OoII group mainly applies to clusters not very
metal-poor and with very blue HB morphology ([Fe/H]$\ge -$2.2 and HB
type $\ge$ +0.8, filled squares) since the $k_{puls}$ values typical
of the variables observed in clusters with very low metal abundance
and moderately blue HB morphology (e.g., M15-like) seem to be
absent. The lack of this kind of variables shows up quite clearly
from Fig. 8 which shows the frequency distribution of the
$k(1.5)_{puls}$ values in $\omega$~Cen (bottom) in comparison with
those for OoI and OoII clusters.  Note that this result, which holds
also if the new metal abundances by Sollima et al. (2006) are
adopted, cannot be explained by invoking a significant difference between the
Kraft \& Ivans (2003) and the Rey et al. (2000) metallicity scales. 
By using the Gratton et al. (2004) metal
abundance [Fe/H]$_G$ determinations for RR Lyrae stars in NGC~1851,
NGC~3201, and in NGC~4590 we get [Fe/H]$_G\sim
-$0.48+0.65[Fe/H]$_K$, while for $\omega$~Cen variables we derive
[Fe/H]$_G\sim -$0.41+0.71[Fe/H]$_R$. Eventually, we find  
[Fe/H]$_K\sim$ 0.1+1.1[Fe/H]$_R$.

\begin{figure}
\begin{center}
\includegraphics[width=8cm]{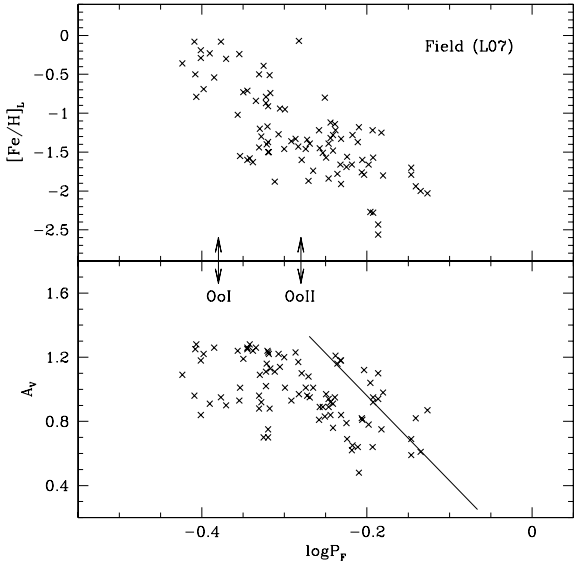}
\caption{
RR$_{ab}$ stars in the solar neighborhood with measured [Fe/H] abundances
(top panel; Layden 2007, private communication) and visual amplitudes
(bottom panel; Nikolov, Buchantsova \& Frolov 1984). The arrows mark the
shortest period observed in OoI and OoII Galactic globular clusters. The
solid line shows the predicted ridge line for cluster OoII variables
(see Fig. 3).
}
\end{center}
\end{figure}

\begin{figure}
\begin{center}
\includegraphics[width=8cm]{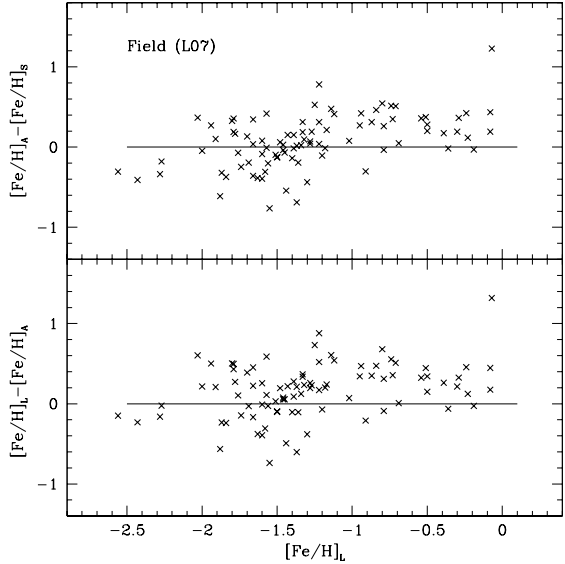}
\caption{Difference between the observed and the calculated [Fe/H] 
versus period for all the stars in Fig. 9.}
\end{center}
\end{figure}

\subsection{Field RR$_{ab}$ stars}

Figure 9 shows the $PA_V$ diagram of RR$_{ab}$ stars in the solar 
neighborhood for which [Fe/H] (Layden 1995,1998,2007, hereinafter [L07], private 
communication) and $A_V$ data (Nikolov, Buchantsova \& Frolov 1984) are 
available. These stars are a mixture of OoII and OoI variables, with a 
further population at shorter periods than the OoI limit (see also the 
analysis of Kinemuchi et al. (2006) of a quite huge sample of field 
variables.) and [Fe/H]$\sim -$0.5.   
As a first test, we show in Fig. 10 the difference between the measured 
metal content [Fe/H]$_L$ and the calculated values [Fe/H]$_A$ from Eq. (1) 
and [Fe/H]$_S$ from Eq. (2). In both cases, the average difference is 
$\sim \pm$ 0.3 dex, but the discrepancy for individual variables 
may be two or three times larger.

\begin{figure}
\begin{center}
\includegraphics[width=8cm]{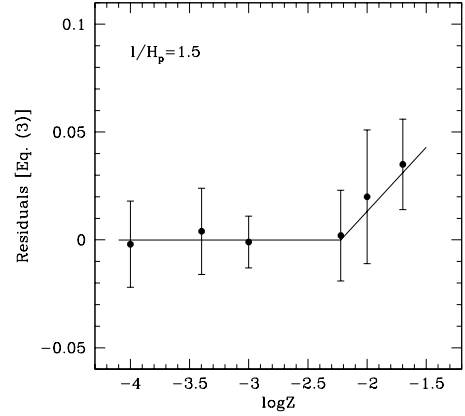}
\caption{Residuals to Eq. (3) for all the fundamental 
models from $Z$=0.0001 to $Z$=0.02.}
\end{center}
\end{figure}

\begin{figure}
\begin{center}
\includegraphics[width=8cm]{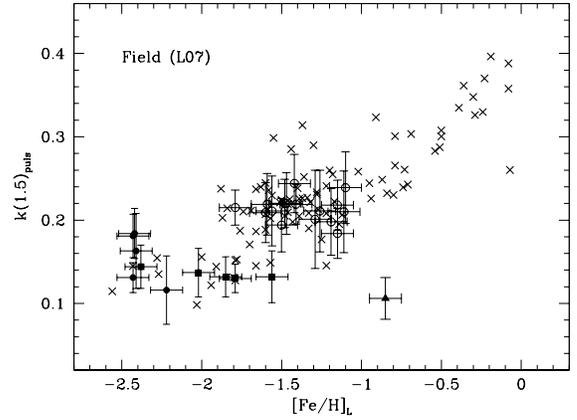}
\caption{Pulsational parameter  $k(1.5)_{puls}$ as a function of metal abundance 
[Fe/H]$_L$ for all the field RR$_{ab}$ stars plotted in Fig. 9. The average values 
for Galactic globular clusters (symbols as in Fig. 5) and NGC~6441 (triangle) 
are also reported.} 
\end{center}
\end{figure}

To repeat the procedure adopted for the variables in Galactic globular clusters, 
we have first verified that Eqs. (3) and (4) hold for fundamental RR Lyrae stars 
with $Z >$0.006. As shown in Fig. 11, our pulsation models constructed by adopting 
$l/H_p$=1.5 and $Z$=0.01, 0.02 (Bono et al. 1997) suggest that the constant 
term in Eq. (3) changes as 0.136+0.06(log$Z$+2.22). On this ground, we determine 
the $k(1.5)_{puls}$ values plotted in Fig. 12. 
A glance at the data plotted in this figure discloses three relevant points:

\begin{itemize}
\item the field stars, at constant $k_{puls}$, show a dispersion in 
iron abundance that might reach $\pm$ 0.5 dex.
\item  The $k_{puls}$ parameter steadly decreases, on average, when moving 
from [Fe/H]$\sim$ 0 to [Fe/H]$\sim -$2, but without any further decrease 
for the most metal-poor ([Fe/H]$\sim -$2.4) variables;
\item  The behavior of field stars and Galactic globular clusters, in 
the metallicity range [Fe/H]=$-$1.0 to $-$2.5, is quite similar. 
This finding supports the evidence for similar physical and evolutionary  
properties for field and cluster variables within this metallicity range.
\item There is a significant difference between the RR$_{ab}$ 
variables in NGC~6441 and field variables with similar metal content. 
   
\end{itemize}

\section{Exploiting the $PA_V$ diagram}
\subsection{Period-Amplitude-Magnitude relation for RR$_{ab}$ stars}
The circumstantial empirical and theoretical evidence
discussed in the above sections brought into focus the
deceptive use of the Bailey diagram of RR$_{ab}$ stars
to estimate metal abundances. Therefore we are now facing
the question: is there any possibility to exploit its dependence
on the evolutionary status of the variables?

It is well known that current updated HB models provide, for fixed
helium and metal content, slightly different luminosity values which
are due to different assumptions on input physics (see, e.g.,
Castellani 2003). On the contrary, the predicted mass of the RR Lyrae
stars appears a more safe parameter, with an average variation of
$\sim$ 2\% among the various evolutionary prescriptions available
in the recent literature. On this ground, it has already been shown in
Paper II and Paper III that the coupling between the predicted
relations inferred by the pulsation models, where mass and luminosity 
are free parameters, and the pulsator average mass
suggested by SHB simulations provides a reliable ``pulsational''
route to the determination of the absolute magnitude of RR Lyrae
stars {\it in globular clusters with known metal content and HB
morphology}.

\begin{figure}
\begin{center}
\includegraphics[width=8cm]{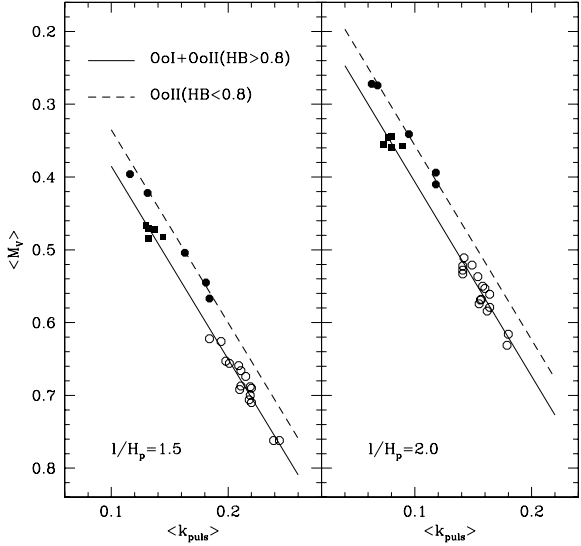}
\caption{{\it Left panel:} Mean absolute magnitude of RR$_{ab}$ stars
in Galactic globular clusters versus $\langle k_{puls}\rangle$ for $l/H_p$=1.5
and  by adopting scaled-solar chemical compositions and the solar ratio
$(Z/X)_{\odot}$=0.0245. The symbols are the same as in Fig. 5. The solid line
is Eq. (5), while the dashed lines has a brighter zero-point (0.05 mag).
{\it Right panel:} Same as in the left panel, but with 
$l/H_p$=2.0. The solid line is Eq. (6).}
\end{center}
\end{figure}

\begin{table*}
\begin{center}
\caption{Average mass $M$(RR) of RR$_{ab}$ stars in Galactic
globular clusters inferred by SHB computations, adopting
solar-scaled chemical compositions and $(Z/X)_{\odot}$=0.0245. These
masses are used with Eq. (2) and Eq. (3) to estimate the 
visual distance moduli $\langle
\mu_V\rangle$ and the mean absolute magnitudes
$\langle M_V\rangle$ isted in columns (5)-(8).} \label{PLCM}
\begin{tabular}{lccccccc}
\hline \hline
Name&HB&[Fe/H]$_K$&$M$(RR)&$\mu_V^{k(1.5)}$& $\langle M_V^{k(1.5)}\rangle$ &$\mu_V^{k(2.0)}$&$\langle M_V^{k(2.0)}\rangle$\\	
\hline\\																	
\multicolumn{8}{c}{Oosterhoff	type	II}\\															
N4590	&   +0.44& $-$2.43&	0.81&	15.06$\pm$0.09& 0.54$\pm$0.13& 15.21$\pm$0.09&	 0.39$\pm$0.13	\\	
N6426	&   +0.58& $-$2.43&	0.80&	17.75$\pm$0.07& 0.42$\pm$0.11& 17.90$\pm$0.06&	 0.27$\pm$0.11	\\	
N7078	&   +0.67& $-$2.42&	0.78&	15.23$\pm$0.08& 0.57$\pm$0.10& 15.38$\pm$0.08&	 0.41$\pm$0.10	\\	
N5053	&   +0.52& $-$2.41&	0.81&	16.07$\pm$0.10& 0.50$\pm$0.11& 16.23$\pm$0.10&	 0.34$\pm$0.11	\\	
N6341	&   +0.91& $-$2.38&	0.72&	14.59$\pm$0.09& 0.48$\pm$0.11& 14.71$\pm$0.08&	 0.36$\pm$0.11	\\	
N5466	&   +0.58& $-$2.22&	0.77&	16.06$\pm$0.09& 0.40$\pm$0.11& 16.19$\pm$0.07&	 0.27$\pm$0.11	\\	
N5024	&   +0.81& $-$2.02&	0.71&	16.32$\pm$0.07& 0.47$\pm$0.10& 16.45$\pm$0.06&	 0.34$\pm$0.10	\\	
N6809	&   +0.87& $-$1.85&	0.69&	13.85$\pm$0.07& 0.47$\pm$0.11& 13.98$\pm$0.05&	 0.35$\pm$0.11	\\	
N6333	&   +0.87& $-$1.79&	0.68&	15.72$\pm$0.06& 0.47$\pm$0.10& 15.82$\pm$0.05&	 0.36$\pm$0.10	\\	
N7089	&   +0.96& $-$1.56&	0.66&	15.46$\pm$0.09& 0.48$\pm$0.13& 15.59$\pm$0.09&	 0.36$\pm$0.13	\\	
\hline\\																	
\multicolumn{8}{c}{Oosterhoff	type	I}\\								
N4147	&   +0.55& $-$1.79&	0.71&	16.24$\pm$0.11& 0.67$\pm$0.14& 16.35$\pm$0.11&	0.56$\pm$0.14	\\	
I4499	&   +0.11& $-$1.60&	0.70&	16.99$\pm$0.08& 0.66$\pm$0.10& 17.13$\pm$0.07&	0.52$\pm$0.10	\\	
N6934	&   +0.25& $-$1.59&	0.70&	16.19$\pm$0.12& 0.69$\pm$0.12& 16.32$\pm$0.10&	0.55$\pm$0.12	\\	
N3201	&   +0.08& $-$1.56&	0.70&	14.08$\pm$0.13& 0.67$\pm$0.13& 14.21$\pm$0.12&	0.54$\pm$0.12	\\	
N5272	&   +0.08& $-$1.50&	0.69&	15.00$\pm$0.08& 0.63$\pm$0.10& 15.12$\pm$0.07&	0.51$\pm$0.10	\\	
N7006	& $-$0.28& $-$1.48&	0.70&	18.11$\pm$0.14& 0.69$\pm$0.15& 18.24$\pm$0.14&	0.55$\pm$0.15	\\	
N6715	&   +0.75& $-$1.47&	0.67&	17.38$\pm$0.08& 0.71$\pm$0.11& 17.52$\pm$0.09&	0.57$\pm$0.11	\\	
N6981	&   +0.14& $-$1.42&	0.68&	16.10$\pm$0.11& 0.76$\pm$0.15& 16.24$\pm$0.11&	0.62$\pm$0.15	\\	
N6229	&   +0.24& $-$1.41&	0.68&	17.35$\pm$0.10& 0.70$\pm$0.13& 17.47$\pm$0.10&	0.58$\pm$0.13	\\	
N6864	& $-$0.07& $-$1.29&	0.67&	17.02$\pm$0.15& 0.66$\pm$0.16& 17.15$\pm$0.18&	0.52$\pm$0.16	\\	
N5904	&   +0.31& $-$1.26&	0.66&	14.38$\pm$0.09& 0.69$\pm$0.13& 14.50$\pm$0.08&	0.57$\pm$0.13	\\	
N1851	& $-$0.36& $-$1.19&	0.66&	15.40$\pm$0.12& 0.65$\pm$0.12& 15.52$\pm$0.11&	0.53$\pm$0.12	\\	
N6121	& $-$0.06& $-$1.15&	0.66&	12.73$\pm$0.06& 0.62$\pm$0.10& 12.82$\pm$0.05&	0.53$\pm$0.10	\\	
N6362	& $-$0.58& $-$1.15&	0.66&	14.56$\pm$0.06& 0.71$\pm$0.10& 14.68$\pm$0.05&	0.58$\pm$0.10	\\	
N6723	& $-$0.08& $-$1.11&	0.65&	14.68$\pm$0.13& 0.69$\pm$0.16& 14.80$\pm$0.13&	0.57$\pm$0.16	\\	
N6171	& $-$0.73& $-$1.10&	0.65&	14.91$\pm$0.14& 0.76$\pm$0.15& 15.04$\pm$0.13&	0.63$\pm$0.15	\\	
\hline
\end{tabular}
\end{center}
\end{table*}

Then, we estimate the average mass of RR Lyrae stars in the selected
globular clusters using the SHBs listed in Table 2, under the
hypothesis of scaled-solar chemical compositions. In order to transform
the measured [Fe/H] value into the global metallicity $Z$, we adopt the
solar value $(Z/X)_{\odot}$=0.0245 (Grevesse \& Noels 1993) and $f$=1
in the relation log$Z$=[Fe/H]$-$1.73+log(0.638$f$+0.362), 
where $f$ is the enhancement factor of $\alpha$-elements with respect 
to iron (Salaris et al. 1993). The predicted mass values, which have 
an intrinsic uncertainty of $\sim$ 2\%, are
listed in column (4) of Table 5 and, once inserted into Eqs. (3) and
(4), they provide the visual distance moduli $\mu_V^{k(1.5)}$ and
$\mu_V^{k(2.0)}$ and the RR$_{ab}$ mean absolute magnitudes
$\langle M_V^{k(1.5)}\rangle$ and $\langle M_V^{k(2.0}\rangle$
given in columns (5)-(8) in the same Table.

Data plotted in Fig. 13, where the error bars are not drawn for 
the sake of clearness, show the direct consequence of the
HB morphology-metallicity progression disclosed in Fig. 2:  
the RR$_{ab}$ stars
observed in OoII clusters with HB type bluer than +0.8 (filled
squares) and in OoI clusters (open circles) obey a common 
relations between the absolute magnitude and the $k_{puls}$ parameter, 
as given by 

$$\langle M_V^{k(1.5)}\rangle=0.12(\pm0.10)+2.65(\pm0.07)\langle k(1.5)_{puls}\rangle\eqno(5)$$
and
$$\langle M_V^{k(2.0)}\rangle=0.14(\pm0.10)+2.67(\pm0.07)\langle k(2.0)_{puls}\rangle,\eqno(6)$$
\noindent
while for the RR Lyrae variables in OoII clusters with moderately
blue HB morphology (filled circles) the zero-points of the above 
relations (dashed lines) are moderately brighter by $\sim$ 0.05 mag.

\begin{table*}
\begin{center}
\caption{Selected results of SHB simulations with $Z\ge$ 0.002. 
For each given metal abundance, we list the mean mass of HB stars 
producing very blue and very red HB types and 
the corresponding mean mass of RR Lyrae stars. 
The last column gives the average mass (logarithm) 
of the predicted RR Lyrae stars for the whole range 
from HB=+0.95 to HB=$-$0.95. All the mass values 
hold for old stellar structures (see text).}\label{PLCM}
\begin{tabular}{lcccccc}
\hline \hline

$Z$&[Fe/H]&$\langle M$(HB)$\rangle$&$\langle M$(RR)$\rangle$&$\langle M$(HB)$\rangle$&
$\langle M$(RR)$\rangle$&$\langle$log$M$(RR)$\rangle$\\
 &  & HB=+0.95 & HB=+0.95& HB=$-$0.95 & HB=$-$0.95 & \\ 
\hline																					
0.002	&$-$0.96& 0.58&0.62$\pm$0.02 & 0.69&0.65$\pm$0.03&$-$0.200$\pm$0.024\\
0.003	&$-$0.79& 0.56&0.60$\pm$0.02 & 0.66&0.62$\pm$0.03&$-$0.217$\pm$0.021\\
0.004	&$-$0.66& 0.55&0.59$\pm$0.02 & 0.65&0.61$\pm$0.02&$-$0.225$\pm$0.013\\
0.006	&$-$0.49& 0.54&0.58$\pm$0.02 & 0.64&0.59$\pm$0.02&$-$0.234$\pm$0.010\\
0.008	&$-$0.37& 0.53&0.56$\pm$0.01 & 0.63&0.58$\pm$0.01&$-$0.245$\pm$0.008\\
0.01	&$-$0.27& 0.52&0.56$\pm$0.01 & 0.62&0.57$\pm$0.01&$-$0.249$\pm$0.007\\
0.02	&$+$0.03& 0.51&0.54$\pm$0.01 & 0.58&0.55$\pm$0.01&$-$0.264$\pm$0.005\\
\hline
\end{tabular}
\end{center}
\end{table*}

Regarding the field RR Lyrae stars, we do not know the morphology of the 
parent HB star distribution, but luckily enough we can take benefit by the 
well-known evidence that, for a fixed age, the predicted 
mass range of HB stars populating the RR Lyrae instability strip decreases with 
increasing the metal content. This is shown in Table 6, where the data  
already presented in Table 2 are implemented by new SHB results at $Y$=0.25 
(Caputo et al. 2007, in preparation) based on Pietrinferni et al. (2004, 2006) 
HB models produced by RGB progenitor having an age of about 13 Gyr. 
Adopting [Fe/H]=1.73+log$Z$, a linear regression through 
the average values listed in the last column in this Table gives 

$$\langle \log M(RR)\rangle=-0.265-0.063[Fe/H],\eqno(7)$$ 
\noindent
with the intrinsic uncertainty given by 
$\epsilon (\langle$log$M$(RR)$\rangle)=0.005-0.02$[Fe/H]. 
According to Eq. (3) and bearing in mind that with larger 
metal content than $Z$=0.006 the constant term varies as 
0.136+0.06(log$Z$+2.22), we eventually derive 
that the absolute magnitude of RR$_{ab}$ stars 
is given by
$$M_V^{k(1.5)}=0.56-0.49A_V-2.60\log P+0.05[Fe/H]\eqno(8)$$ 
with $-1.0\le$[Fe/H]$\le -$0.5 and by 
$$M_V^{k(1.5)}=0.64-0.49A_V-2.60\log P+0.20[Fe/H]\eqno(9)$$
with $-0.5\le$[Fe/H]$\le$0, 
with the magnitude total uncertainty varying as 
$\epsilon (M_V)$=0.07$-$0.02[Fe/H].

\subsection{$M_V$(RR)-[Fe/H] relation}

Many calibrations of the RR Lyrae luminosity as 
function of the metal content have been published in the 
relevant literature (e.g., see Cacciari \& Clementini 2003 
for a summary) and the most recent ones suggest that the $M_V$(RR)-[Fe/H]  
is nonlinear for metal abundances ranging from [Fe/H]$\sim -$0.5 to 
$-$2.4 (see Sandage 2006; Sandage \& Tammann 2006, and 
references therein).

\begin{figure}
\begin{center}
\includegraphics[width=8cm]{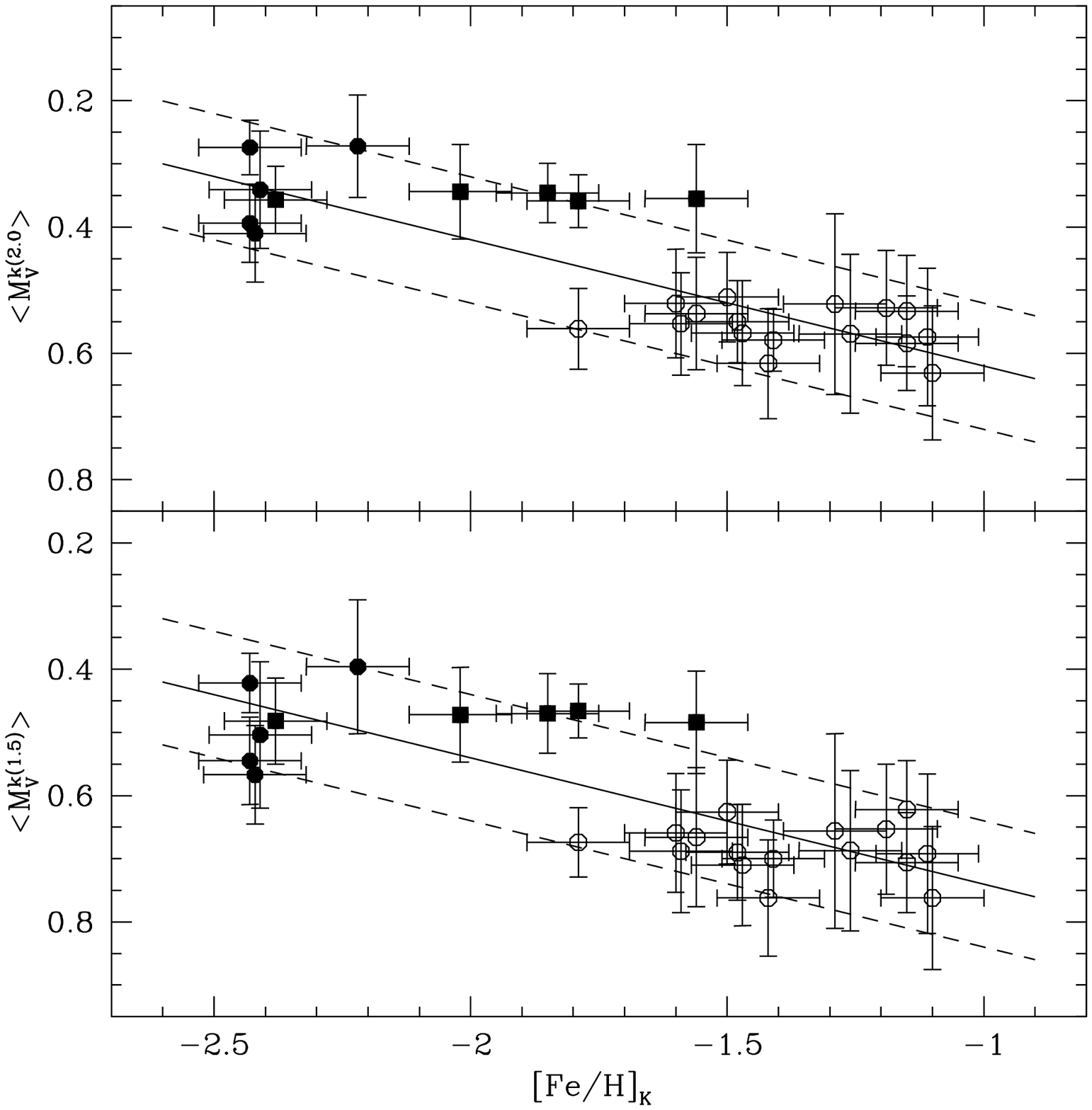}
\caption{Mean absolute visual magnitudes $\langle M_V^{k(1.5)}\rangle$ 
(bottom panel) and $\langle M_V^{k(2.0}\rangle$ (top panel) of 
RR$_{ab}$ stars in Galactic globular clusters versus [Fe/H]$_K$, 
according to scaled-solar chemical compositions and a solar ratio 
$(Z/X)_{\odot}$=0.0245.
The symbols are the same as in Fig. 5. The solid line shows the linear
regression through the entire sample and has a slope of 0.20 mag 
dex$^{-1}$, while the dashed lines the 1$\sigma$ uncertainty. See text 
for more details.}
\end{center}
\end{figure}

\begin{figure}
\begin{center}
\includegraphics[width=8cm]{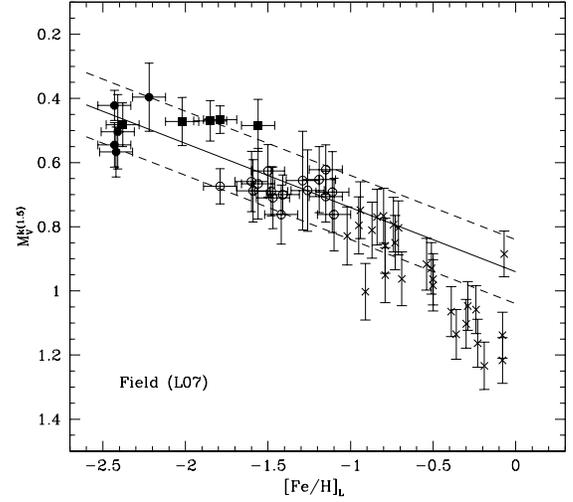}
\caption{Absolute visual magnitudes $M_V^{k(1.5)}$ versus [Fe/H]$_L$ for 
field RR$_{ab}$ stars more metal-rich than [Fe/H]$_L=-$1.0 in comparison with 
Galactic globular cluster variables. Symbols and lines are the same 
as in the bottom panel of Fig. 14.}
\end{center}
\end{figure}

\begin{figure}
\begin{center}
\includegraphics[width=8cm]{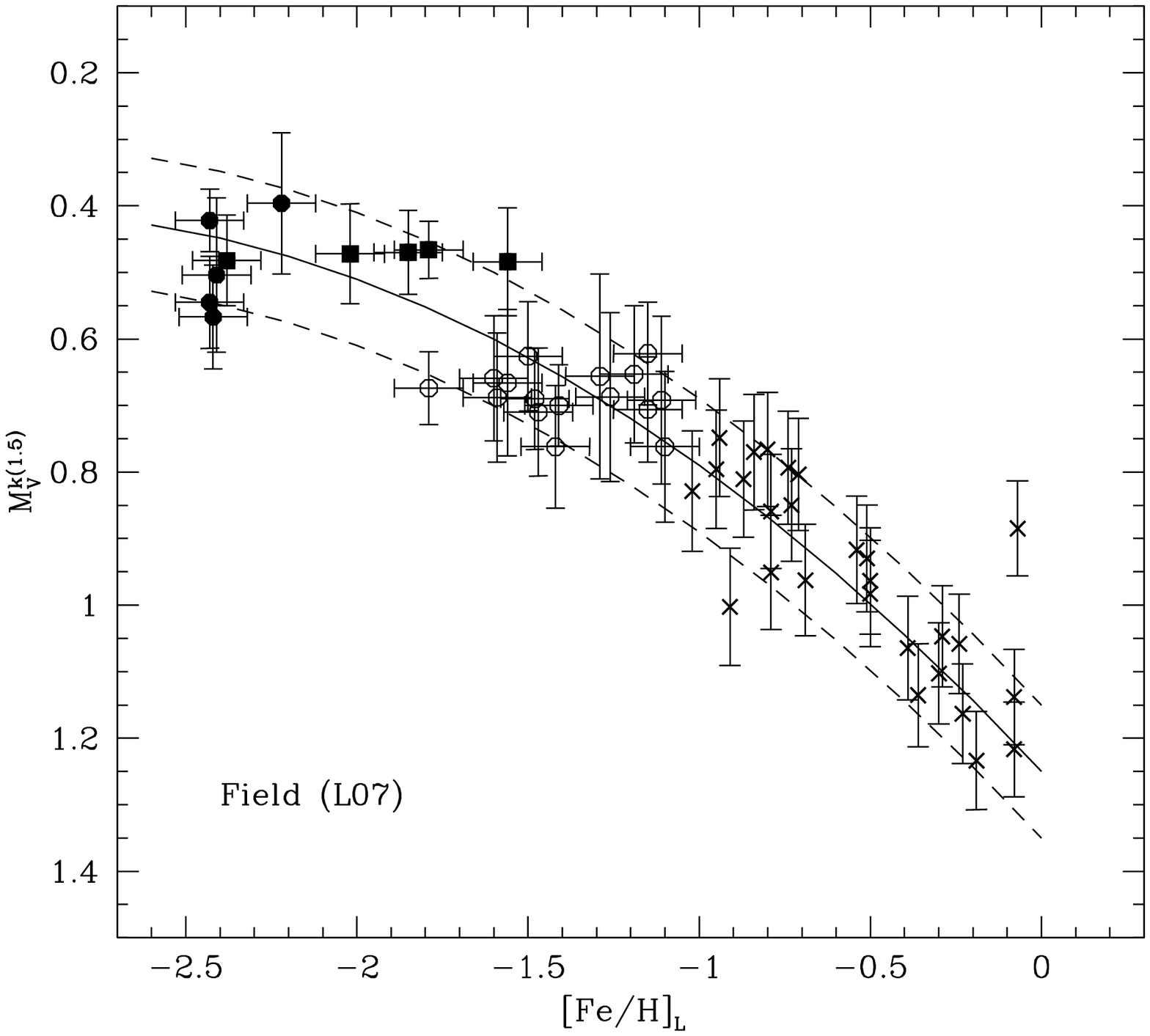}
\caption{Same as Fig. 14, but with cluster and field data fitted with a 
quadratic relation.} 
\end{center}
\end{figure}

For our selected sample of Galactic globular clusters, 
Fig. 14 displays the $PA_V$-based mean absolute magnitude of 
RR$_{ab}$ stars (columns (6) and (8) in Table 5)   
versus the
cluster metallicity [Fe/H]$_K$. The linear regression over the
entire sample (solid line) yields a slope of 0.20$\pm$0.06 mag
dex$^{-1}$, regardless of the adopted mixing-length parameter, 
while the zero-point of the relation changes 
from 0.94$\pm$0.10 mag to 0.82$\pm$0.10 
mag with $l/H_p$=1.5 and 2.0, respectively. However, the data given 
in Table 5 clearly shows that at constant metal content the RR $_{ab}$ 
luminosity depends on the cluster HB type: e.g., the variables in 
NGC~7089 (HB=+0.96) are $\sim$ 0.2 mag brighter than those in 
IC~4499, NGC~6934 and NGC~3201, which show a HB type from HB=+0.08 to 
+0.25, 
yet all these clusters have nearly the same metallicity. 
This result is not a novelty 
since theoretical (see Paper IV and references therein) and 
observational studies (Lee \& Carney 1999; Clement \& Shelton 1999; 
Alves, Bond \& Onken 2001) have already suggested that the RR Lyrae absolute 
magnitude depends on the cluster HB morphology and metal content. 

The comparison with field RR$_{ab}$ stars  
with [Fe/H]$\ge -$1.0 is shown in Fig. 15, where the absolute magnitudes 
of the field variables are 
determined by using Eqs. (8) and (9). It is quite clear that the linear 
$M_V$(RR)-[Fe/H] relation provided by Galactic globular clusters is not 
suitable to the most metal-rich ([Fe/H]$\ge -$0.7) field variables. Conversely, 
we show in Fig. 16 that over the whole metallicity range of 
[Fe/H]=$-$2.5 to $\sim$ 0 {\it all} the variables are nicely fitted 
by the quadratic relation 
$$M_V^{k(1.5)}=1.19(\pm0.10)+0.50[Fe/H]+0.09[Fe/H]^2\eqno(10)$$

\begin{figure}
\begin{center}
\includegraphics[width=8cm]{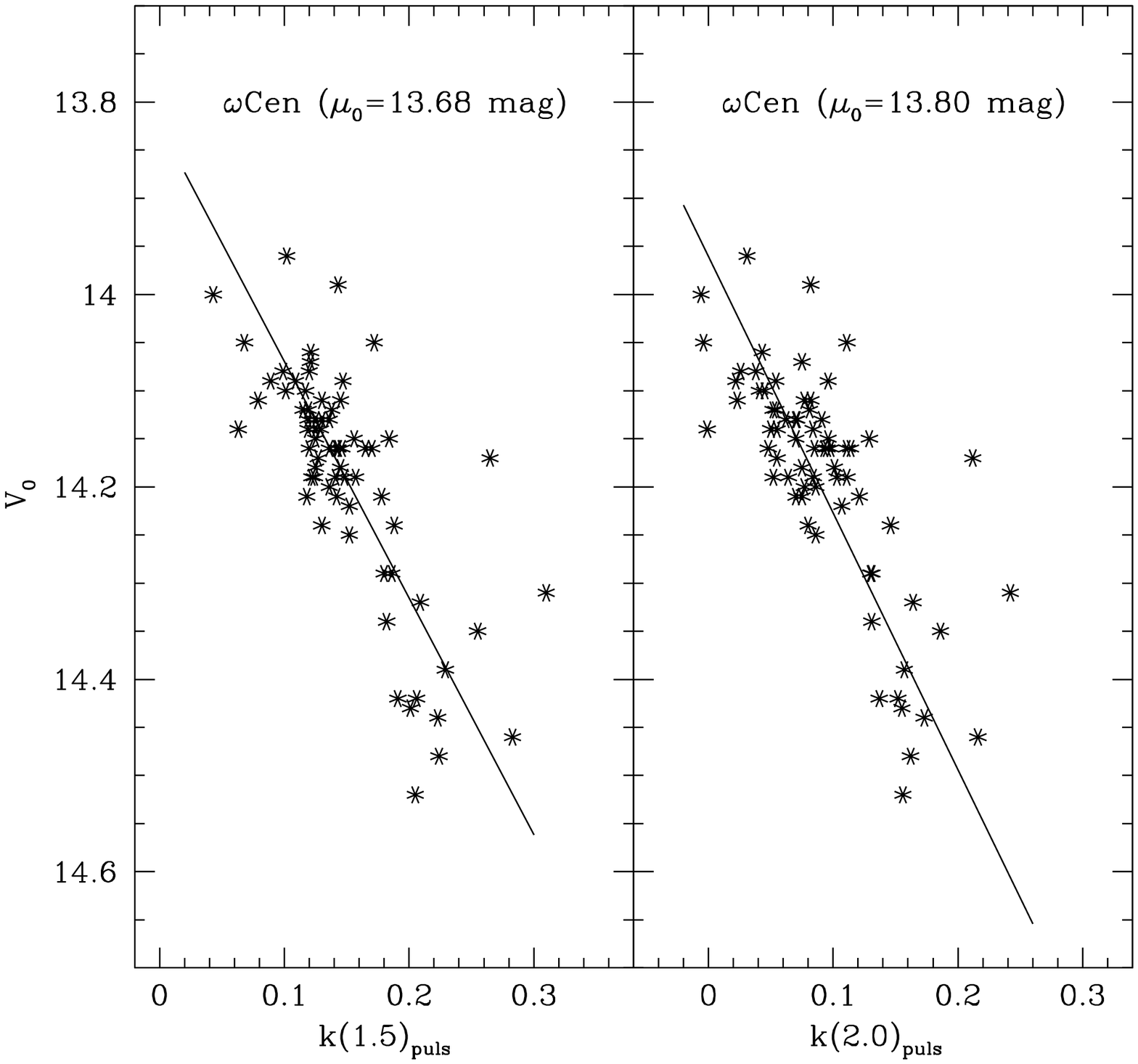}
\caption{Unreddened visual magnitude $V_0$ of RR$_{ab}$ stars
in $\omega$~Cen plotted versus $k(1.5)_{puls}$ and $k(2.0)_{puls}$.
The solid line plotted in the left panel refers to Eq. (5) and accounts
for an intrinsic distance modulus $\mu_0$=13.68 mag. The solid line
plotted in the right panel refers to Eq. (6) and accounts for 
$\mu_0$=13.80 mag.}
\end{center}
\end{figure}

\begin{figure}
\begin{center}
\includegraphics[width=8cm]{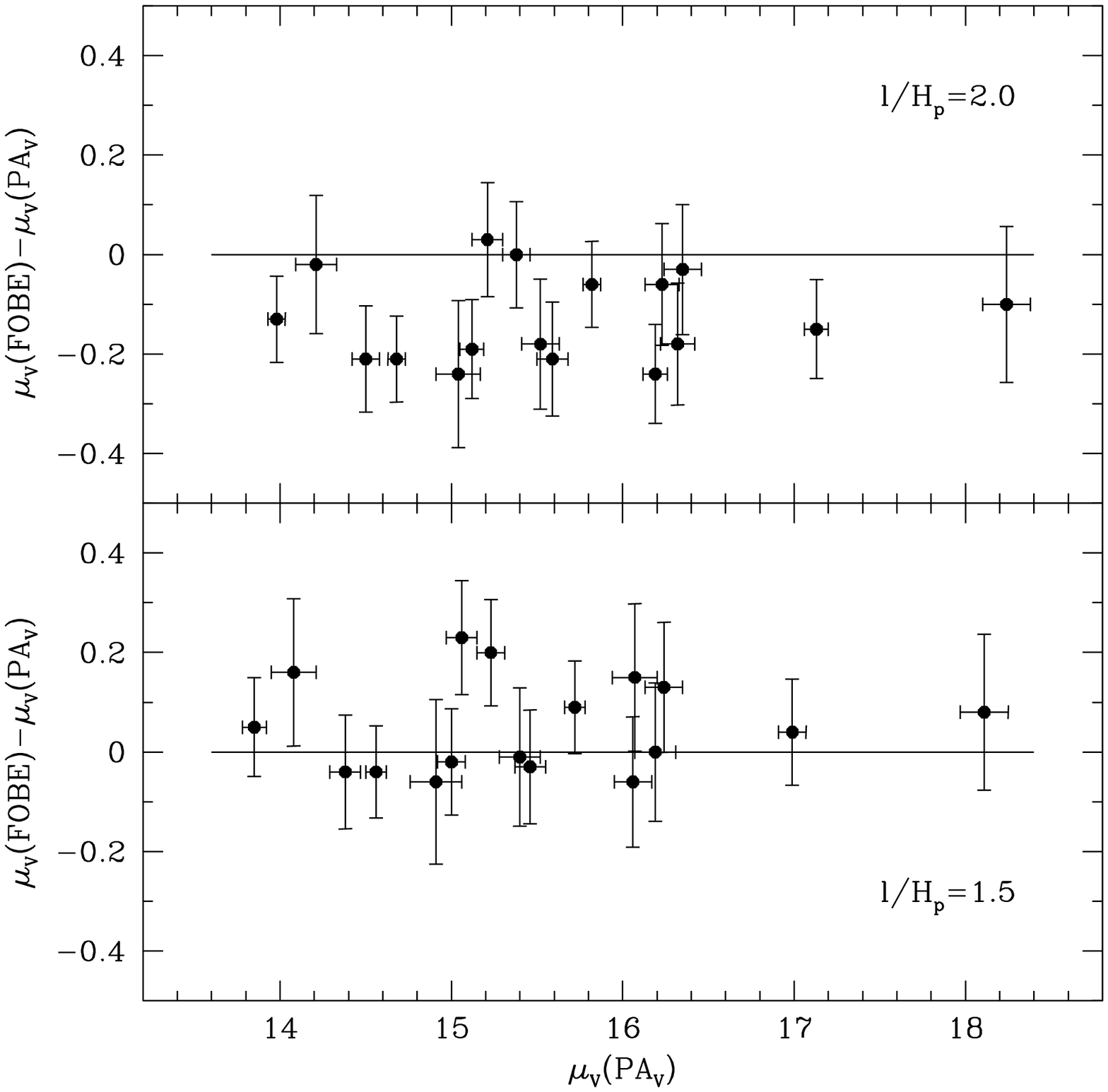}
\caption{Comparison between the apparent distance moduli of RR$_c$
variables based on the FOBE method and the RR$_{ab}$ distance moduli based
on the $PA_V$ relation for the two adopted values of the mixing-length
parameter. The data refer to scaled-solar chemical compositions and
to the solar ratio $(Z/X)_{\odot}$=0.0245.}
\end{center}
\end{figure}

\subsection{Which value of the mixing length parameter?}

We have shown that the value of the mixing length parameter 
influences the zero-point of the Period-Amplitude-Magnitude 
relation [Eqs. (5) and (6)] and consequently the 
$M_V$(RR)-[Fe/H] calibration (see Fig. 14).
    
In order to constrain the most appropriate value of the
mixing-length parameter for globular cluster RR$_{ab}$ 
stars, we show in Fig. 17 the $V_0$ magnitudes of
RR$_{ab}$ stars in $\omega$~Cen (Piersimoni et al. 2007, in
preparation) versus the observed $k(1.5)_{puls}$ and $k(2.0)_{puls}$
parameters. By using Eqs. (5) and (6), we find a cluster intrinsic
distance modulus of $\mu_0$=13.68$\pm$0.10 mag and 13.80$\pm$0.10
mag, respectively. Unfortunately, both these estimates agree within
1$\sigma$ with the distance $\mu_0$=13.75$\pm0.04$ mag based on the
eclipsing binary OGLEGC-17 (Thompson et al. 2001; Kaluzny et al.
2002). Therefore, we decided to consider a further pulsational
method, namely the FOBE method (Caputo 1997; 
Caputo et al. 2000) which provides the cluster apparent distance modulus by matching the
observed distribution of the RR$_c$ variables in the $V$-log$P$
plane with the predicted blue (hot) edge of 
the first-overtone instability region. 
The reason of this choice deals with the fact that the FOBE-based
distance modulus $\mu_V$(FOBE) is expected to decrease with
increasing the mixing-length parameter (see Eq. (2) in Paper III),
at variance with the apparent distance $\mu_V(PA_V)$ inferred from
the $PA_V$ relation. 

Figure 18 shows the comparison between the two
sets of distance moduli. We find that for  $l/H_p$=1.5 the  
$\mu_V$(FOBE) distances are on average larger than those 
based on $\mu_V(PA_V)$, whereas the opposite applies for 
$l/H_p$=2.0. This evidence indicates that we can adopt
$l/H_p\sim$ 1.7, although the best solution discussed in 
Paper III is probably given by a mixing-length parameter 
that slightly increases  when moving from the blue to the 
red side of the instability strip, 
i.e., from $c$- to $ab$-type variables. In this context, it is worth 
noting that the very recent investigation by Ferraro et al. (2006) 
on red giant stars in globular clusters supports a value $l/H_p$=2.0 
for these cool stars and a negligible dependence on metallicity.

Finally, we note that the use of different scalings between the iron
abundance and the global metallicity ($Z$-[Fe/H]) has marginal
effects on the RR$_{ab}$ absolute magnitudes listed in Table 5. As a
matter of fact, by adopting $f$=3 ([$\alpha$/Fe]$\sim$ 0.5) with
$(Z/X)_{\odot}$=0.0245 yields, at fixed [Fe/H], smaller masses by
$\sim$ 6\%, and in turn fainter absolute magnitudes by $\sim$ 0.03
mag, when compared with the values listed in Table 5. The dependence
on the adopted solar ratio is even smaller, and indeed by adopting
$(Z/X)_{\odot}$=0.0165 (Asplund et al. 2004), the mass and magnitude
variations for $f$=1 are only $\sim$ +3\% and $-$0.01 mag, while for
$f$=3 we estimate $\sim -$3\% and $\sim$+0.01 mag, respectively.

%%%%%%%%%%%%%%%%%%%%%%%%%%%%%%%%%%%%%%%%%%%%%%%%%%%%%%%%%%%%%%%%%%%%%%%%%%
\section{Conclusions and final remarks}

Hydrodynamical models of fundamental RR Lyrae stars computed by adopting 
metal content from $Z$=0.0001 to 0.006 and   
two different values of the mixing-length parameter ($l/H_p$=1.5 and 2.0) 
provide detailed predictions concerning the pulsation parameters connecting 
the period with the $V$-band amplitude. In order to investigate the 
distribution of cluster RR$_{ab}$ stars in the $PA_V$ diagram, we consider 
the following pulsational parameters  

$$k(1.5)_{puls}=0.13-\log P_{ab}-0.189A_V$$
and 
$$k(2.0)_{puls}=0.03-\log P_{ab}-0.142A_V,$$
\noindent 
and we find that the average values $\langle k(1.5)_{puls}\rangle$ 
and $\langle k(2.0)_{puls}\rangle$ do not show significant changes 
among OoI clusters with metal abundances ranging from [Fe/H]=$-$1.8 
to $-$1.1 and intermediate to red HB types. 
On the other hand, the same parameters present a mild decrease among 
the OoII clusters with very blue HB types, even if these clusters are 
also the less metal-poor of the group. 
Moreover, in the relatively narrow metallicity range [Fe/H]=$-1.7\pm$0.1, 
where both OoI and OoII clusters are observed, the former clusters have 
redder HB types {\it and} larger $\langle k_{puls}\rangle$ values than 
the latter ones.

A linear fit over the entire sample of globular clusters yields a 
[Fe/H]-$k_{puls}$ relation with a large intrinsic dispersion of  
$\approx 0.4$ dex. The dispersion becomes even larger if the 
calibration relies on selected clusters: if we adopt a mix  
of OoI and OoII clusters with moderately blue HB morphology, then 
the metal abundance of RR$_{ab}$ in clusters characterized by 
a very blue HB morphology will be underestimated by $\approx 0.7$ dex, whereas 
if we adopt a mix of OoI and OoII clusters with 
very blue HB morphologies the metallicity of RR$_{ab}$ in clusters 
characterized by a moderately blue HB morphology will be overestimated 
by $\approx 0.5$ dex. This circumstantial evidence casts several doubts 
on the use of the $PA_V$ distribution of RR$_{ab}$ variables as a 
diagnostic of the metal abundance. 
This finding is independently supported by the sizable samples of 
RR$_{ab}$ variables in $\omega$ Cen and and in the solar neighborhood 
for which are available metal abundance and $V-$band amplitudes. 
The distribution of these objects in the $PA_V$ plane shows that 
the spread in metal abundance, at constant $k_{puls}$, is of the order 
of 0.5 dex.

By coupling pulsation models and synthetic horizontal branch simulations, 
we show that 
the pulsation parameter $k_{puls}$ is a reliable distance indicator for 
globular clusters with known metal content and HB type. We wish to mention 
that the occurrence of a 
Period-Luminosity-Amplitude relation for RR$_{ab}$ stars was originally 
suggested by Sandage (1981a,b) and that the present use of detailed 
evolutionary and pulsational predictions provides the opportunity to 
constrain the dependence on the globular cluster HB type and metal 
content. On this ground, we 
find that the RR$_{ab}$ in OoI clusters and in OoII clusters with HB types
bluer than +0.8 do obey to well defined $M_V$-$k_{puls}$ relations. 
In particular, we find 
$$\langle M_V^{k(1.5)}\rangle=0.12(\pm0.09)+2.65(\pm0.07)\langle k(1.5)_{puls}\rangle$$
and
$$\langle M_V^{k(2.0)}\rangle=0.14(\pm0.09)+2.67(\pm0.07)\langle k(2.0)_{puls}\rangle,$$
\noindent
while the RR$_{ab}$ in OoII clusters with moderately
blue HB morphology present, at fixed $k_{puls}$, a zero-point that is 
$\sim$ 0.05 mag brighter. 
Regarding the variables in the solar neighborhood, additional 
pulsation models with $l/H_p$=1.5 and 
$Z>$0.006 together with the predicted metallicity dependence of the mass of 
metal-rich ([Fe/H]$\ge-$1.0) RR Lyrae stars    
$$\langle \log M(RR)\rangle=-0.265-0.063[Fe/H]$$
\noindent yield  
$$M_V^{k(1.5)}=0.56-0.49A_V-2.60\log P+0.05[Fe/H]$$ 
with $-1.0\le$[Fe/H]$\le -$0.5 and  
$$M_V^{k(1.5)}=0.64-0.49A_V-2.60\log P+0.20[Fe/H]$$
with $-0.5\le$[Fe/H]$\le$0.  
   
Once the $PA_V$-based absolute magnitude $M_V$(RR) is derived, the 
resulting correlation with the globular cluster metallicity [Fe/H]$_K$ 
has a slope of 0.20$\pm$0.06 mag dex$^{-1}$, 
regardless of the adopted mixing-length parameter, while  
the zero-point changes from 0.94$\pm$0.10 to 0.82$\pm$0.10 mag 
when using pulsation models constructed by assuming a mixing length 
parameter $l/H_p$=1.5 and $l/H_p$=2.0, respectively. However, the inclusion of 
the metal-rich field variables yields that over the total metallicity range 
from [Fe/H]=$-$2.5 to $\sim$ the relation becomes quadratic as 
$$M_V^{k(1.5)}=1.19(\pm0.10)+0.50[Fe/H]+0.09[Fe/H]^2$$
\noindent 
in agreement with the results presented by by 
Bono et al. (2003) and Sandage (2006).  

Finally, in order to constrain the most appropriate value 
of the mixing-length parameter, we adopt the RR$_{ab}$ stars 
in $\omega$~Cen, but the $PA_V$-based true distance moduli, 
$\mu_0$=13.68$\pm$0.09 mag for $l/H_p$=1.5 and 13.80$\pm$0.09 mag 
for $l/H_p$=2.0, agree within 1$\sigma$ with the distance 
$\mu_0$=13.75$\pm0.04$ mag based on the eclipsing binary 
OGLEGC-17 (Thompson et al. 2001; Kaluzny et al.  2002). 
Therefore, we adopt the FOBE method that provides cluster 
apparent distance moduli which decrease with increasing the 
mixing-length parameter. Eventually, we find that distance 
estimates based on the $PA_V$ and on the FOBE method 
agree for an intermediate mixing-length parameter, namely 
$l/H_p\sim$ 1.7.

\begin{acknowledgements}
It is a real pleasure to thank H. Smith for several suggestions 
and a detailed reading of an early draft of this paper. We also 
warmly thank A. Layden for his valuable data on field RR Lyrae 
stars and his helpful comments.  
We also acknowledge the anonymous referee for his/her positive 
comments and suggestions that helped us to improve the readability 
of the manuscript. 
This work was partially supported by PRIN-INAF2005 (P.I.: A. Buzzoni),
"Galactic Stellar Populations", by PRIN-INAF2004 (P.I.: M. Bellazzini),
"A hierarchical merging tale told by stars: motions, ages and chemical
compositions within structures and substructures of the Milky Way".
\end{acknowledgements}

%%%%%%%%%%%%%%%%%%%%%%%%%%%%%%%%%%%%%%%%%%%%%%%%%%%%%%%%%%%%%%%%%%%%%%%%%%%%%%%%%%%
%\begin{references}
%%%%%%%%%%%%%%%%%%%%%%%%%%%%%%%%%%%%%%%%%%%%%%%%%%%%%%%%%%%%%%%%%%%%%%%%%%%%%%%%%%
%\pagebreak 

\end{document}